\documentclass[twocolumn]{aastex63}

\usepackage{lipsum, babel}
\usepackage{lineno}

\newcommand{\be}{\begin{equation}}
\newcommand{\ee}{\end{equation}}
\newcommand{\bex}{\begin{equation}\notag}
\newcommand{\eex}{\end{equation}\notag}
\newcommand{\bea}{\begin{eqnarray}}
\newcommand{\eea}{\end{eqnarray}}
\newcommand{\beax}{\begin{eqnarray*}}
\newcommand{\eeax}{\end{eqnarray*}}
\newcommand{\ba}{\begin{array}}
\newcommand{\ea}{\end{array}}

\received{}
\revised{}
\accepted{}
\submitjournal{ApJ}

\shorttitle{Statistics of Photospheric Magnetic Cancellations}
\shortauthors{Ledvina et al.}

\begin{document}

\title{Quantifying Properties of Photospheric Magnetic Cancellations in the Quiet Sun Internetwork}

\correspondingauthor{Vincent E. Ledvina}
\email{vincent.ledvina@und.edu}

\author[0000-0003-0127-5105]{Vincent E. Ledvina}
\affiliation{University of North Dakota}
\affil{National Solar Observatory, 3665 Discovery Dr., Boulder, CO 80303, USA}

\author[0000-0001-8975-7605]{Maria D. Kazachenko}
\affil{National Solar Observatory, 3665 Discovery Dr., Boulder, CO 80303, USA}

\author[0000-0002-4525-9038]{Serena Criscuoli}
\affil{National Solar Observatory, 3665 Discovery Dr., Boulder, CO 80303, USA}

\author{Dennis Tilipman}
\affiliation{University of Colorado}

\author[0000-0003-2596-9523]{Ilaria Ermolli}
\affil{INAF – Osservatorio Astronomico di Roma, Via Frascati 33, Monte Porzio Catone, RM, 00078, Italy}

\author[0000-0002-0018-6488]{Mariachiara Falco}
\affil{INAF - Catania Astrophysical Observatory, Via Santa Sofia 78, I–95123 Catania, Italy}

\author[0000-0002-1837-2262]{Salvatore Guglielmino}
\affil{INAF - Catania Astrophysical Observatory, Via Santa Sofia 78, I–95123 Catania, Italy}

\author[0000-0002-7711-5397]{Shahin Jafarzadeh}
\affil{Max Planck Institute for Solar System Research, Justus-von-Liebig-Weg 3, 37077 G\"{o}ttingen, Germany}
\affil{Rosseland Centre for Solar Physics, University of Oslo, P.O. Box 1029 Blindern, NO-0315 Oslo, Norway}

\author[0000-0003-2088-028X]{Luc Rouppe van der Voort}
\affil{Rosseland Centre for Solar Physics, University of Oslo, P.O. Box 1029 Blindern, NO-0315 Oslo, Norway}
\affil{Institute of Theoretical Astrophysics, University of Oslo, P.O. Box 1029 Blindern, NO-0315 Oslo, Norway}

\author[0000-0003-1853-2550]{Francesca Zuccarello}
\affil{Dipartimento di Fisica e Astronomia “Ettore Majorana”—Sezione Astrofisica, Università di Catania, Via S. Sofia 78, I-95123 Catania, Italy}



\begin{abstract}


We analyzed spectropolarimetric data from the Swedish 1-meter Solar Telescope to investigate physical properties of small-scale magnetic cancellations in the quiet Sun photosphere. Specifically, we looked at the full Stokes polarization profiles along the \ion{Fe}{1} 557.6 nm and of the \ion{Fe}{1} 630.1 nm lines measured by CRisp Imaging SpectroPolarimeter (CRISP) to study temporal evolution of the line-of-sight (LOS) magnetic field during 42.5 minutes of quiet Sun evolution. From this magnetogram sequence, we visually identified 38 cancellation events. We then used Yet Another Feature Tracking Algorithm (YAFTA) to characterize physical properties of these magnetic cancellations. We found on average $1.6\times10^{16}$ Mx of magnetic flux cancelled in each event with an average cancellation rate of $3.8\times10^{14}$ Mx s$^{-1}$. The derived cancelled flux is associated with strong downflows, with an average speed of $V_\mathrm{LOS}\approx1.1$ km s$^{-1}$. Our results show that the average lifetime of each event is $9.2$ minutes with an average $44.8\%$ of initial magnetic flux being cancelled. Our estimates of magnetic fluxes provide a lower limit since studied magnetic cancellation events have magnetic field values that are very close to the instrument noise level. We observed no horizontal magnetic fields at the cancellation sites and therefore can not conclude whether the events are associated structures that could cause magnetic reconnection.
\end{abstract}




\section{Introduction} \label{sec:intro}
The Sun has historically been divided into two domains: the active Sun and the quiet Sun (QS). The active Sun is commonly defined as areas of the Sun occupied by active regions, plage, and sunspots while the quiet Sun represents the remaining areas. It is reported that in the early stages of Solar Physics, scientists believed the QS to be non-magnetic because only granular convection could be seen in continuum images \citep[e.g.][]{bellotrubio2019}. 
However, early measurements showed that magnetic features are ubiquitous on the Sun. For example, using the Kitt Peak magnetograph, \citet{weaksignals} reported a background internetwork field level of 2-3~G. Recent spectropolarimetric measurements allow magnetic structures to be observed down to scales at the limits imposed by current spatial resolution (e.g. \citet{qsfieldresolved}). These can be analyzed by using new inversion techniques to interpret the Zeeman and Hanle effects (e.g. \citet{editedbib1}). 
Quite the opposite of non-magnetic, the QS displayed a reticular pattern of intense kilogauss fields, the magnetic network (NE), and a varied distribution of smaller-scale (sub-arcsec) magnetic flux concentrations in the areas between them - the solar internetwork \citep[e.g.][]{introduction_7}. State-of-the-art, three-dimensional magnetohydrodynamic simulations of the solar atmosphere \citep[e.g.][]{rempel2014} indicate that large part of the solar surface magnetic features is still unresolved. 

Studies by \citet{Internetwork_1} and others (e.g. \citealt{introduction_1}) have indicated that internetwork (IN) fields are essential contributors to the Sun's overall magnetic flux output, with transport of magnetic flux to the solar photosphere at a rate of $120$ Mx cm$^{-2}$ day$^{-1}$, which is significantly higher than the $1$ Mx cm$^{-2}$ day$^{-1}$ transported by active regions (\citealt{introduction_2}). A large portion of that flux is transported to neighboring intergranular lanes via convective motions (\citealt{introduction_3}) and then to the NE supergranular boundaries (\citealt{introduction_4,introduction_5,introduction_6}). These motions make the IN capable of generating a complete magnetic flux re-supply of the surrounding NE within only $\approx10$ hours (\citealt{introduction_7}) indicating that they are an important contributor to the greater flux output of the solar photosphere. QS  magnetism has in fact been suggested to affect global solar properties such as limb darkening \citep[e.g.][]{criscuoli2017}, photospheric temperature gradient \citep[e.g.][]{faurobert2016} and global radiative output \citep[e.g.][and references therein]{rempel2020}.

The transient nature of the IN manifests in frequent instances of magnetic flux emergence, dissipation, and cancellation events.  IN magnetic flux cancellation is one of three processes (flux decay, cancellation, and interaction with network) in which flux is removed from the photosphere, and is a mechanism that leads to the maintenance of the flux budget in the photosphere \citep{evidence1, introduction_10, Internetwork_1}. A physical cancellation event results in an in-situ disappearance of magnetic flux from the solar photosphere as a result of the interactions between two opposite-polarity magnetic elements \citep{introduction_8,introduction_9}. Cancellations are a contributor to QS magnetic behavior and have been observed to play a critical role in many dynamic upper-atmosphere solar phenomena, such as coronal mass ejections, flares, and filament eruptions \citep{Chintzoglou2019,introduction_11,introduction_12,introduction_13, zuca} as well as the formation of prominences \citep{prominences}, coronal jets \citep{coronaljets}, and Ellerman bombs \citep{ellermanbombs}.

IN cancellations may also partially drive the heating of the chromosphere. \citet{Chromosphere_1} identified $51$ cancellation events using data from the Swedish 1-meter Solar Telescope \citep[SST,][]{scharmer2003} and compared these with chromospheric temperature diagnostics using IRIS data \citep{IRIS}. Magnetic cancellations were found to release enough energy to provide local brightening in the chromosphere \citep{Chromosphere_1_Support_1,Chromosphere_1_Support_2}.



The derivation of statistical properties of IN cancellations is among the most important steps in further understanding their role in other solar phenomena. This is somewhat challenging due to the relatively low signal-to-noise ratio in current polarimetric measurements of QS. IN fields are particularly difficult to observe because they are arranged on small spatial scales, evolve rapidly, and individually, produce very weak signals. Thus, major advances in this research area have mainly been brought upon by advances in observing technologies (instruments with higher spectropolarimetric sensitivity and spatial and temporal resolutions) and new computational modeling, meaning that many of the fundamental aspects of IN cancellations have been discovered recently and are still widely disputed and incomplete. 

In this paper, we analyzed a small area of the QS observed at the SST with high spatial resolution and high cadence spectropolarimetric data. In these data, we noted numerous signatures of magnetic cancellation events, and we describe in the following their statistical properties. Our results complement a growing list of publications focusing on QS magnetism and the important process of cancellations that pervade its surface \citep[e.g.][]{CancellationRates,downflows_omegaloop,Cancellation_1,Cancellation_3,rmsvelocity}

The outline of this paper is as follows: In Section \ref{sec:meth} we describe SST observations and our analysis in Section \ref{sec:data_analysis}. In Section \ref{sec:analysis_of_roi_03} we describe a single cancellation event in detail. In Section \ref{sec:res_stat} we summarized statistical parameters of the physical quantities estimated for all 38 cancellation events. In Section \ref{sec:dis&conc} we discussed our findings and their implications in the broader field of QS research. In Appendix \ref{sec:appendix} we describe four additional cancellation events in detail.

\section{Observations}
\label{sec:meth}
We employed the full Stokes polarization profiles along the Fe I 557.6 nm and the Fe I 630.1 nm lines to analyze the temporal evolution of the magnetic field and the dynamic properties of plasma at cancellation sites.

Specifically, for our analysis we used data acquired at the Swedish 1-meter Solar Telescope \citep[SST;][]{scharmer2003} in La Palma, Spain, with the CRisp Imaging SpectroPolarimeter \citep[CRISP;][]{CRISP}, as part of a two weeks long campaign (August 6-18, 2011). These 
observations captured a quiet-Sun region at disk center on August 6, 2011, beginning at 07:57:39 UTC and lasting for 42.5 minutes.


CRISP acquired data along the \ion{Fe}{1} 630.1, 630.2 and 557.6 nm lines spectral ranges. Only data acquired along the 630.1 and 557.6 nm lines were used in this study. The pixel scale was $\approx$ 0\farcs059/pixel with a field-of-view of $57.5'' \times 57.3''$. A total of 100 scans were acquired with a temporal cadence of 28 s. 
 Raw data were calibrated using an early version of the standard CRISP calibration pipeline \citep[CRISPRED, ][]{CRISPRED}. The SST adaptive optics system \citep{adaptiveoptics} was able to make continuous corrections to the wavefront, effectively operating in a 100\% lock rate. The combination of the use of adaptive optics and of the application of the Multi-Object Multi-Frame Blind Deconvolution image restoration technique \citep[MOMFBD,][]{MOMFBD} allowed for effective minimization of seeing induced aberrations. The images studied here had an angular resolution of $\approx$ $0.15''$ at 557.6 nm, which is close to the diffraction-limit of the SST. The estimated spectropolarimetric sensitivity was $\approx 3 \times 10^{-3}$.


This dataset has been employed in previous studies to investigate the dynamics \citep{kinkwaves1,kinkwaves2} and thermal properties \citep{irradiance, salvocitation} of plasma in the quiet Sun regions. These studies were made possible in part due to the high resolution of the SST and its instruments, inspiring us to analyze the same data to investigate small magnetic features on the QS surface. 

We derived information about the magnetic properties of the plasma using the four Stokes parameters, I, Q, U and V observed in the \ion{Fe}{1} 630.1 nm line. 
Specifically, we derived the line-of-sight magnetic field $B_\mathrm{LOS}$ from the separation of the centroids of the I+V and I-V signals, estimated with the center-of-gravity method \citep{rees1979,uitenbroek2003}:

\begin{equation}
  B_{LOS}=\frac{\lambda_{+}-\lambda_{-}}{2}\frac{4\pi mc}{eg_{L}\lambda_0^2},
\end{equation}
where $g_{L}$ was the line’s
effective Land\'e factor, m and e were the electron mass
and charge, respectively, $\lambda_0$ was the central wavelength of the line, and $\lambda_{\pm}$ were defined as:
\begin{equation}
  \lambda_{\pm}=\frac{\int \lambda(I_c-(I\pm V)) d\lambda  }  {\int (I_c-(I\pm V)) d\lambda },
\end{equation}

where $I_c$ was the line's nearby continuum intensity.
The Total Circular Polarization signal \citep[TCP,][]{deltoro2007} was computed as:

\begin{equation}
  TCP = \int_{\lambda} \frac{V(\lambda)}{I_c}d{\lambda}
\end{equation},

and the Total Linear Polarization signal \citep[TLP,][]{deltoro2007} as:
\begin{equation}
  TLP = \int_{\lambda}\frac{\sqrt{Q(\lambda)^2+U(\lambda)^2)}}{I_c}d{\lambda}
\end{equation}

To estimate the line-of-sight velocities ($V_\mathrm{LOS}$), we used the Doppler shift of the core of the magnetically-insensitive \ion{Fe}{1} 557.6 nm line. The line core was estimated by fitting the observed line profiles with a Gaussian function, and the Doppler shift was computed taking as reference the core position of the average line profile computed over the whole time-series. The resulting average velocity over the whole field of view is -0.07 $\pm$ 0.035 km~s$^{-1}$. Indeed, this value is around "convective upflow" value for the line found in \citet{salvocitation2}. This is approximately -0.15 km~s$^{-1}$ to -0.2 km~s$^{-1}$ at disk center.

\begin{figure*}[tbh!]
 \centering 

\resizebox{0.4\hsize}{!}{\includegraphics[angle=0]{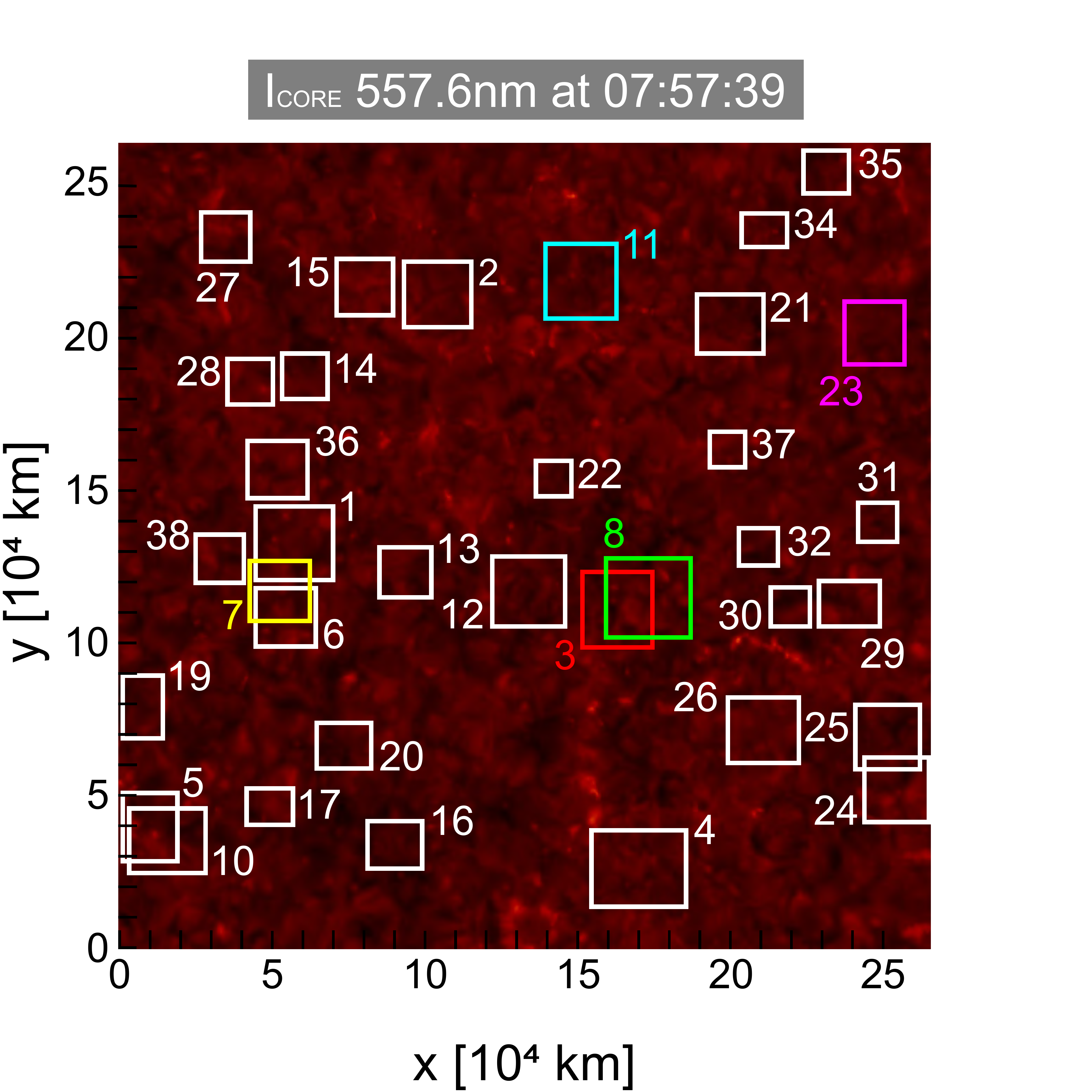}} 
\resizebox{0.4\hsize}{!}{\includegraphics[angle=0]{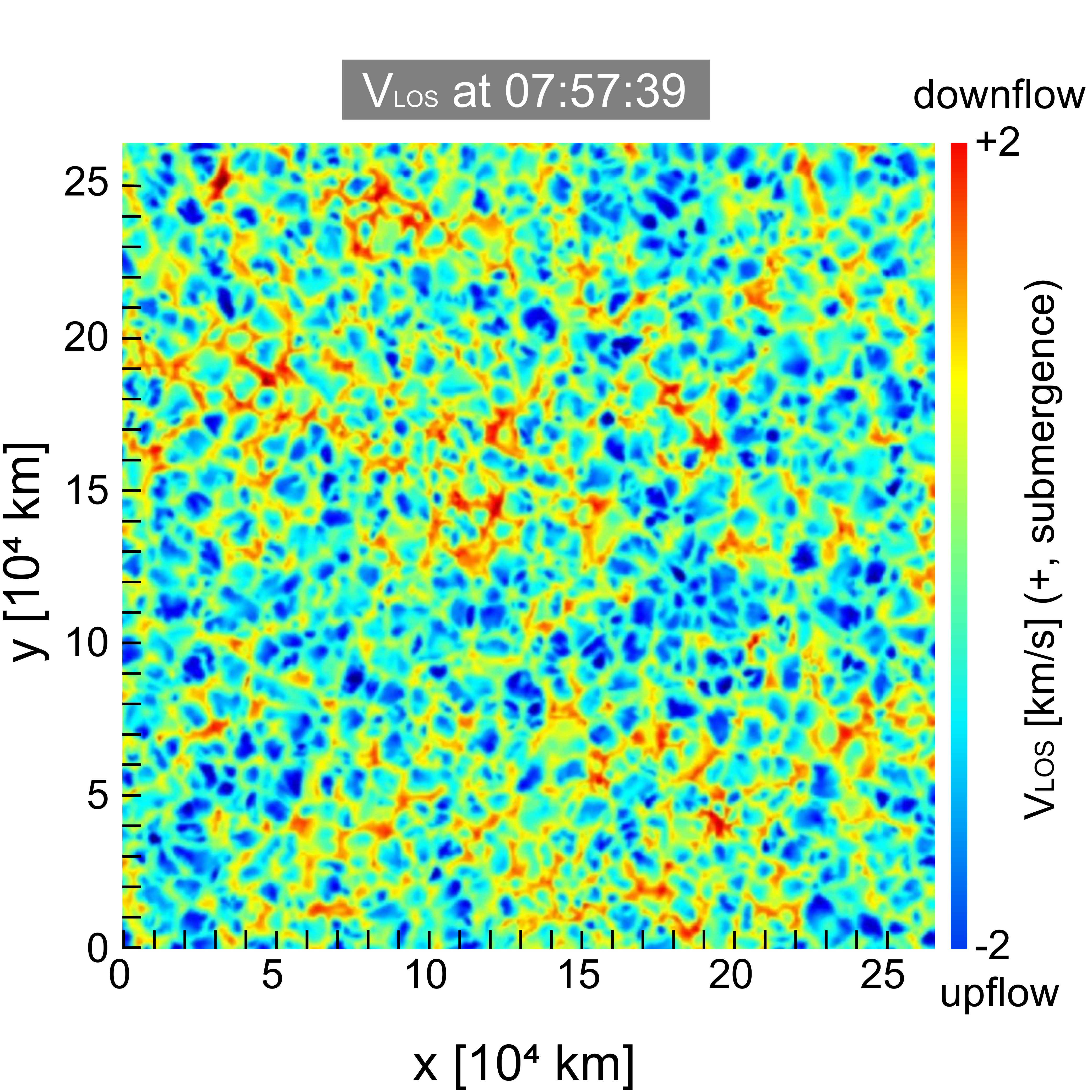}} 
\resizebox{0.4\hsize}{!}{\includegraphics[angle=0]{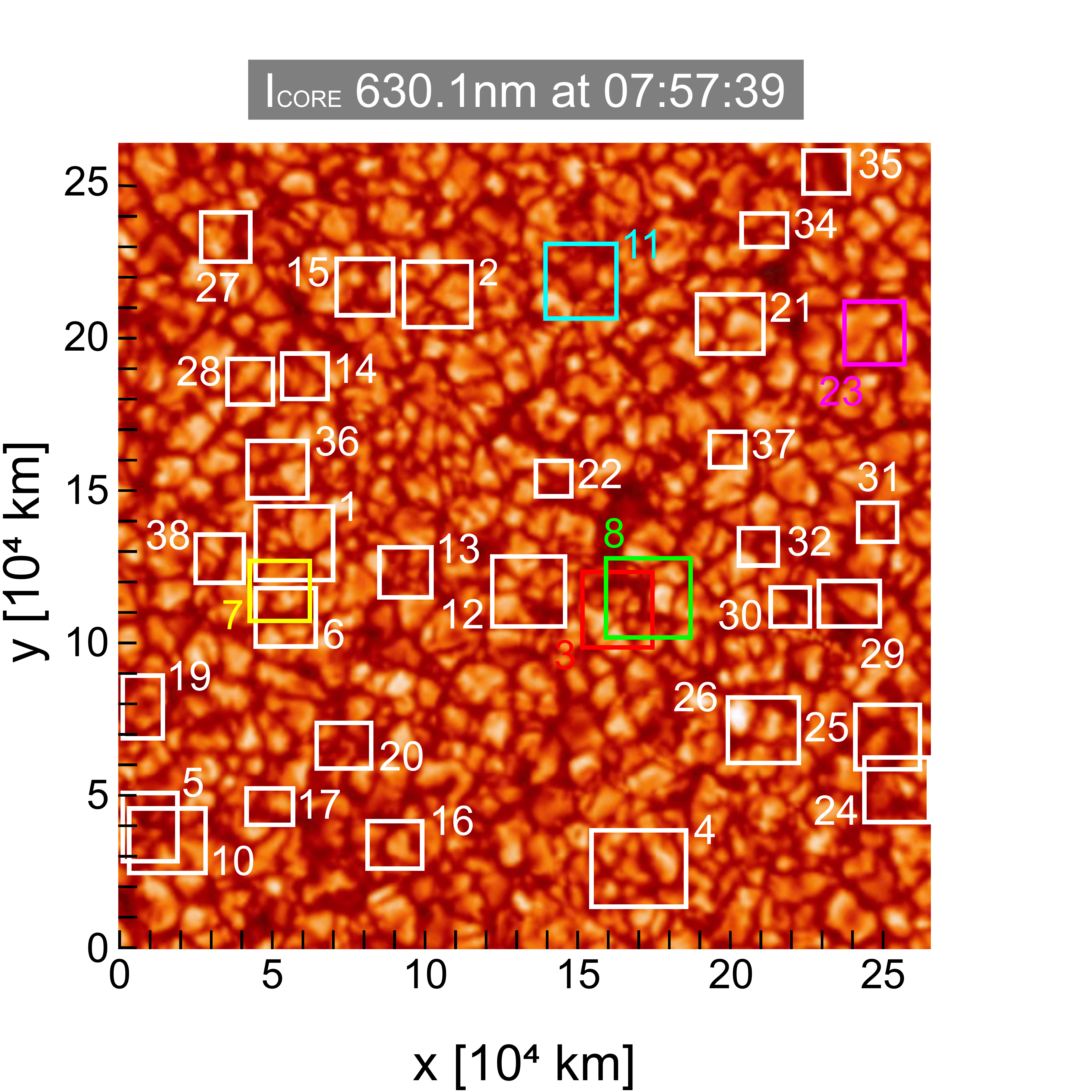}} 
\resizebox{0.4\hsize}{!}{\includegraphics[angle=0]{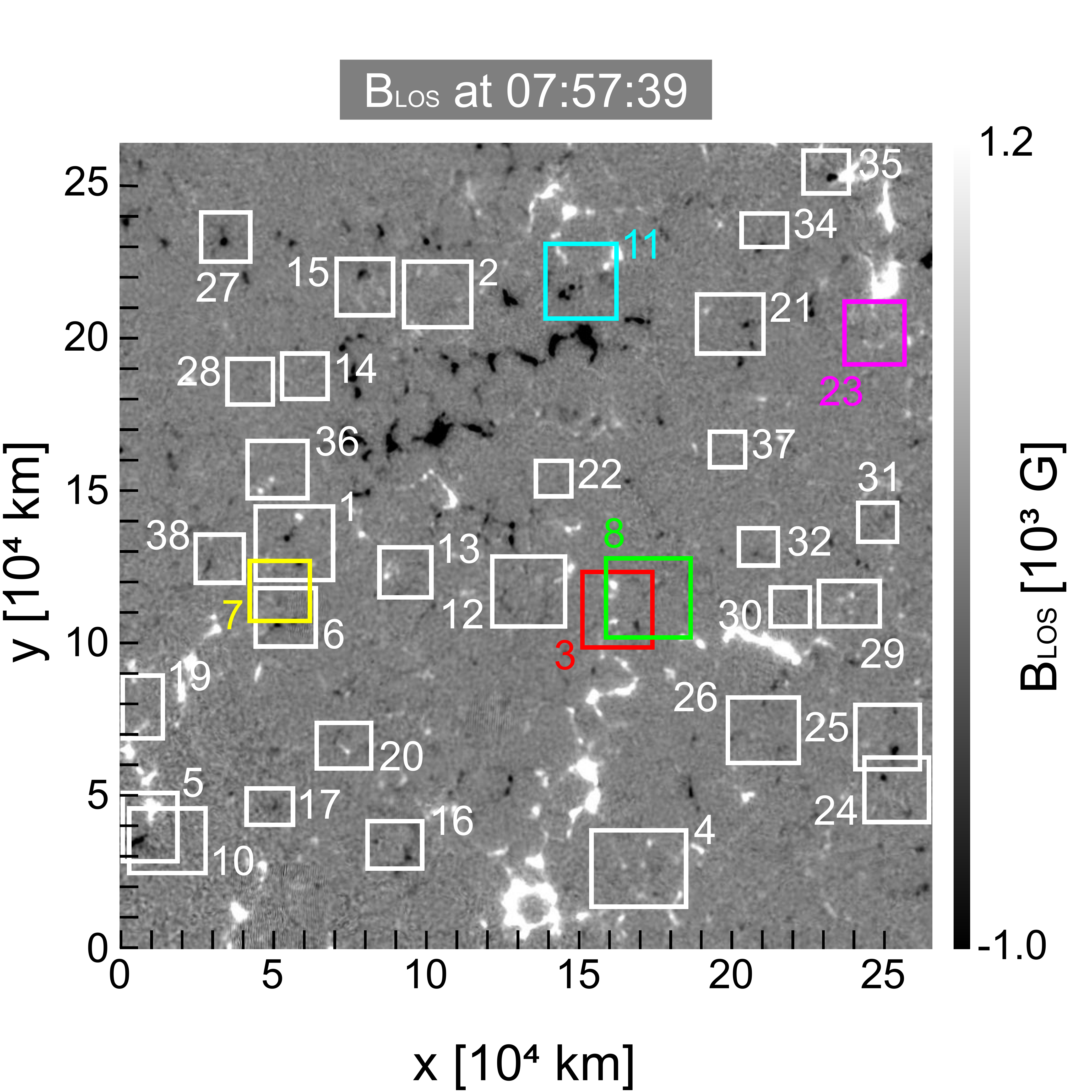}}
 \caption{Four snapshot views of a quiet Sun region at the disk center, as observed by the SST. \textit{Top left}: Core intensity in the non-magnetic \ion{Fe}{1} 557.6 nm line. \textit{Top right}: Line-of-sight velocity ($V_\mathrm{LOS}$) with positive and negative values corresponding to downflows and upflows, respectively. \textit{Bottom left}: Continuum image at \ion{Fe}{1} 630.1 nm. \textit{Bottom right}: Line-of-sight magnetogram (B$_\mathrm{LOS}$). Rectangles indicate analyzed regions of interest (ROI), numbered by the ROI indices. Colored rectangles are events described in section \ref{sec:analysis_of_roi_03} and \ref{sec:appendix} of the manuscript. }
 \label{fig:observations}
\end{figure*}

Examples of the data and data-products used in our analysis are shown in Figure~\ref{fig:observations}: the continuum intensities in the \ion{Fe}{1} 557.6 and 630.1 nm continua (left column) and the derived $V_\mathrm{LOS}$ and $B_\mathrm{LOS}$ (right column). 

\section{Data Analysis}
\label{sec:data_analysis}
We used 
the TCP sequence to visually identify 38 cancellation events for our analysis. We defined the following criteria for our selection. First, for visual detection simplicity, we have chosen cancellations that involve two opposite-sign polarities of roughly equal size. While choice of same-size polarities simplifies the process of cancellation search, it leads to an underestimation of canceled magnetic flux.
Second, we avoided selecting events where same-signed flux recombined or emerged in the same location. Finally, we only selected cancellations where the polarities involved were visually identifiable. The average size of a region of interest (ROI) was $\approx$ $2\times10^{5}$ km$^{2}$. Given these criteria, we identified 38 cancellation regions of interest (ROI), shown as squares in Figure~\ref{fig:observations}.

\begin{figure*}[ht!]
\includegraphics[width=1.0\linewidth]{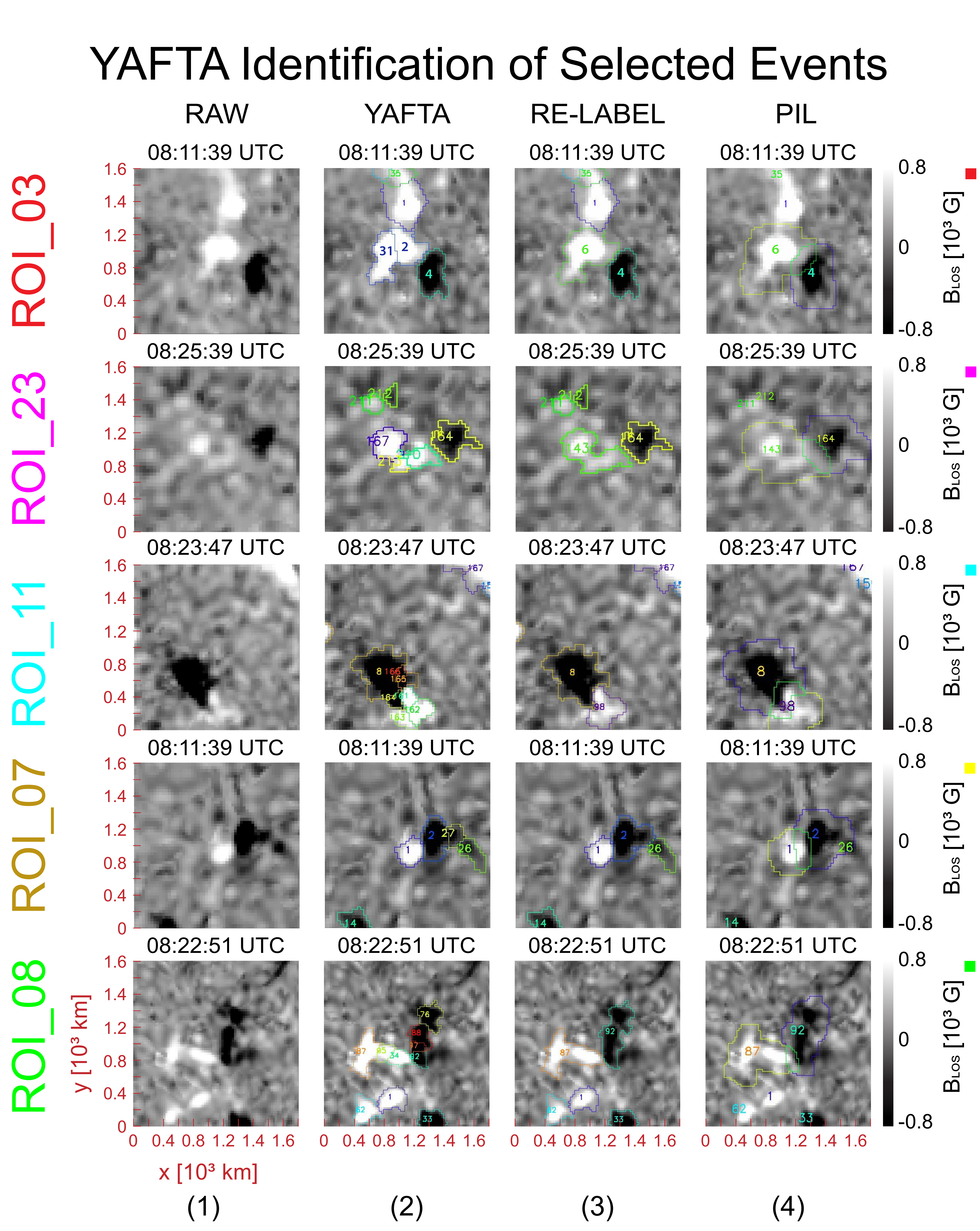}
\caption{B$_\mathrm{LOS}$ time snapshots of $5$ distinct cancellations showing the different steps of processing with YAFTA, as described in Section \ref{sec:meth}. Column (1) shows the raw magnetogram image of the ROI. Column (2) shows the initial tracking result with YAFTA. Column (3) shows the re-labeled features. Column (4) shows the two cancelling features, their dilated masks, and PIL (green outline). Each row of images represents a different ROI. The color of the square beside the right-most image corresponds to the color of the ROI in the full field of view magnetogram in Figure~\ref{fig:observations}. \label{fig:yaftasequence}}
\end{figure*}

For each cancellation identified, we then applied the following three-step process: (1) feature tracking, (2) re-labeling and (3) Polarity Inversion Line (PIL) identification. Examples of the three steps are illustrated in Figure~\ref{fig:yaftasequence} for five events. 
For context, the PIL is the boundary between opposite-polarity features.
In the first step (second column in Figure~\ref{fig:yaftasequence}), to identify individual magnetic features we used "Yet Another Feature Tracking Algorithm" (YAFTA) solar magnetic tracking algorithm \citep{YAFTA_1}. YAFTA has been shown to reliably indentify small and short-lived magnetic features. It identifies magnetic features using a gradient based "downhill" method which dilates local flux maxima by expanding down the gradient toward zero flux density. To discriminate the false-positives, YAFTA allows the user to control the following parameters: a threshold for a minimum magnetic field to consider ($B_\mathrm{min}$), a saddle threshold for a minimum magnetic field to merge the already selected features ($B_\mathrm{saddle}$), and a minimum area size for feature identification ($S_\mathrm{min}$). In our analysis we chose $B_\mathrm{min}=40$ G, $B_\mathrm{saddle}=80$ G, and $S_\mathrm{min}=4$ pixels. 
This parameter set resulted in tracking runs that consistently identified features above the noise level without false-positives. Use of lower thresholds for B$_\mathrm{min}$ resulted in incorrect feature identification in the noisy areas and larger values of feature magnetic fluxes. The thresholds were set based on many tests and visual inspection of results. \citet{threshold} reports the effects of various thresholds used with YAFTA and their implications on feature tracking. After features are identified, YAFTA arbitrarily labels them based on its first pass through the data. In the second step (third column in Figure ~\ref{fig:yaftasequence}), since occasionally YAFTA incorrectly assigned multiple labels to one feature, we had to manually re-label all the features after the initial tracking. This incorrect assignment is due to the fact that YAFTA struggles with very large or strong and very small or weak magnetic elements, reflecting imperfections of our approach. After the second pass the feature masks could be referenced and magnetic properties could be analyzed for each polarity independently. Finally, to describe Doppler velocities associated with each cancellation event, we examined the Doppler velocity properties within the PIL. To define PIL location (fourth column in Figure ~\ref{fig:yaftasequence}), we dilated the masks of the two cancelling polarities by $2$ pixels and defined the PIL as the region where these two masks overlapped. 

To describe each cancellation event we used the following set of parameters: magnetic flux, $\Phi_\mathrm{B}$, flux cancellation rate, $R$, specific cancellation rate, $r$, convergence speed, $V_\mathrm{conv}$, and Doppler velocity, $V_\mathrm{LOS}$. We used these parameters as a foundation for our analysis.

For each observation at time t in the image sequence we defined the unsigned magnetic flux of positive ($+$) and negative ($-$) polarities in each pair, $i$, as 
\begin{equation} \label{eqmag}
 \Phi _{B,i}(t)=\sum_{j=1}^{N_{i}}\left | B_{j}(t) \right | ds^{2} = \left |\Phi_{B,i-} \right |+\left |\Phi_{B,i+} \right |,
\end{equation}
where $N_{i}$ is the number of magnetic field values in the polarity pair, $B_{j}$ is the magnetic field value at each pixel, $j$, and $ds$ is the pixel size (0\farcs059/pixel). 

The position at time $t$ of each positive and negative polarity in the pair was defined as the center of gravity 
$COG_\mathrm{i}(t)=[x_i(t),y_i(t)]$, i.e. the mean position of the feature weighted by the magnetic flux:
\begin{equation} \label{eqxy}
 x_{i}(t)=\frac{\sum_{j=1}^{N_{i}}x_j\left | B_{j} \right |  ds^{2}}{\Phi_{B,i}}, \;\;\;\; 
 y_{i}(t)=\frac{\sum_{j=1}^{N_{i}}y_j\left | B_{j} \right |  ds^{2}}{\Phi_{B,i}}.
\end{equation}
We then used these to calculate the separation distance, $g(t)$, and the convergence velocity, $v_\mathrm{conv}(t)$, between positive and negative polarities in each pair \citep{CancellationRates}:

\begin{equation} \label{eqvel}
  g_i(t)=| COG_{i+}(t)-COG_{i-}(t)|, \\
  v_\mathrm{conv,i}(t)=-0.5\frac{dg_i(t)}{dt}.
\end{equation}


%
To describe the amount of flux cancelled we defined the flux cancellation rate, $R_i$, and the specific cancellation rate (rate per unit PIL length), $\mathrm r_i$:
\begin{equation} \label{eqr}
R_i(t)=\frac{d\Phi_{B,i}}{dt}, \;\;\;\;\;\;\;\; r_i(t)\equiv R_i(t)/l_i(t),
\end{equation}
where $l_i$ was the length of the PIL separating positive and negative polarities of each pair $i$. 

Finally, we found the cancelled flux as the difference between the flux values at the start time ($t_\mathrm{start}$) and end time ($t_\mathrm{end}$) of a cancellation event, so that the positive values indicated cancelled flux: 
\begin{equation} \label{eqvel2}
  \Delta \Phi _{B,i}=\Phi _{B,i}(t_\mathrm{start})-\Phi _{B,i}(t_\mathrm{end}).
\end{equation}
To describe variability of $R_\mathrm i(t)$ over time we also calculated the average and peak cancellation rates, $R_\mathrm{avg}$ and $R_\mathrm{peak}$, respectively, over duration, $T$, of each cancellation event. In the case of complete disappearance of the feature we defined the last frame before disappearance as t$\mathrm{end}$, when the magnetic flux was above zero.

To define the duration, $T$, of each cancellation event, we used the time when the PIL was defined, which occurred when opposite polarities were in close proximity to one another. In the events we observed, while the general trend of the polarities' fluxes was characterized by a decay, sometimes flux momentarily increased during the event, so the PIL was defined in the aforementioned way in order to capture the entire event as a whole.
If, during a cancellation event, one of the polarities dipped below the detection threshold for fewer than $5$ frames but was seen thereafter, we considered it as one cancellation event.

To calculate $V_\mathrm{LOS}$, we averaged the Doppler velocity across the PIL region. Positive values were plasma submergence or downflows, while negative values were emergences or upflows. To describe the $V_\mathrm{LOS}$ variability, we used the mean and the peak values of $V_\mathrm{LOS}$, $V_\mathrm{LOS, avg}$ and $V_\mathrm{LOS, peak}$, respectively. To describe the change in Doppler velocity from the start to the end of the cancellation event we used 
\begin{equation} \label{eqvel3}
  \Delta V_{LOS,i} =V_{LOS,i}(t_\mathrm{start})- V_{LOS,i}(t_\mathrm{end}).
\end{equation}
This formalism allowed us to account for cancellations taking place in intergranular lanes where plasma was already flowing downward or in areas where plasma was already flowing upwards. 

\section{Results}
Figure~\ref{fig:observations} shows a snapshot of the magnetogram sequence that we used to visually identify $38$ distinct cancellation events. Note that we visually inspected the magnetogram to select cancellations in the IN regions away from the strong field regions belonging to the network patches. In this section, we first present our results for one example cancellation event (\S\ref{sec:analysis_of_roi_03}) and then the statistical analysis for all $38$ events (\S\ref{sec:res_stat}). The evolution of an additional four exemplary regions is provided in Appendix~\ref{sec:appendix}. 

\subsection{Example of individual cancellation event, ROI\_03} \label{sec:analysis_of_roi_03}
\begin{figure*}[ht!]
\includegraphics[width=1.0\linewidth]{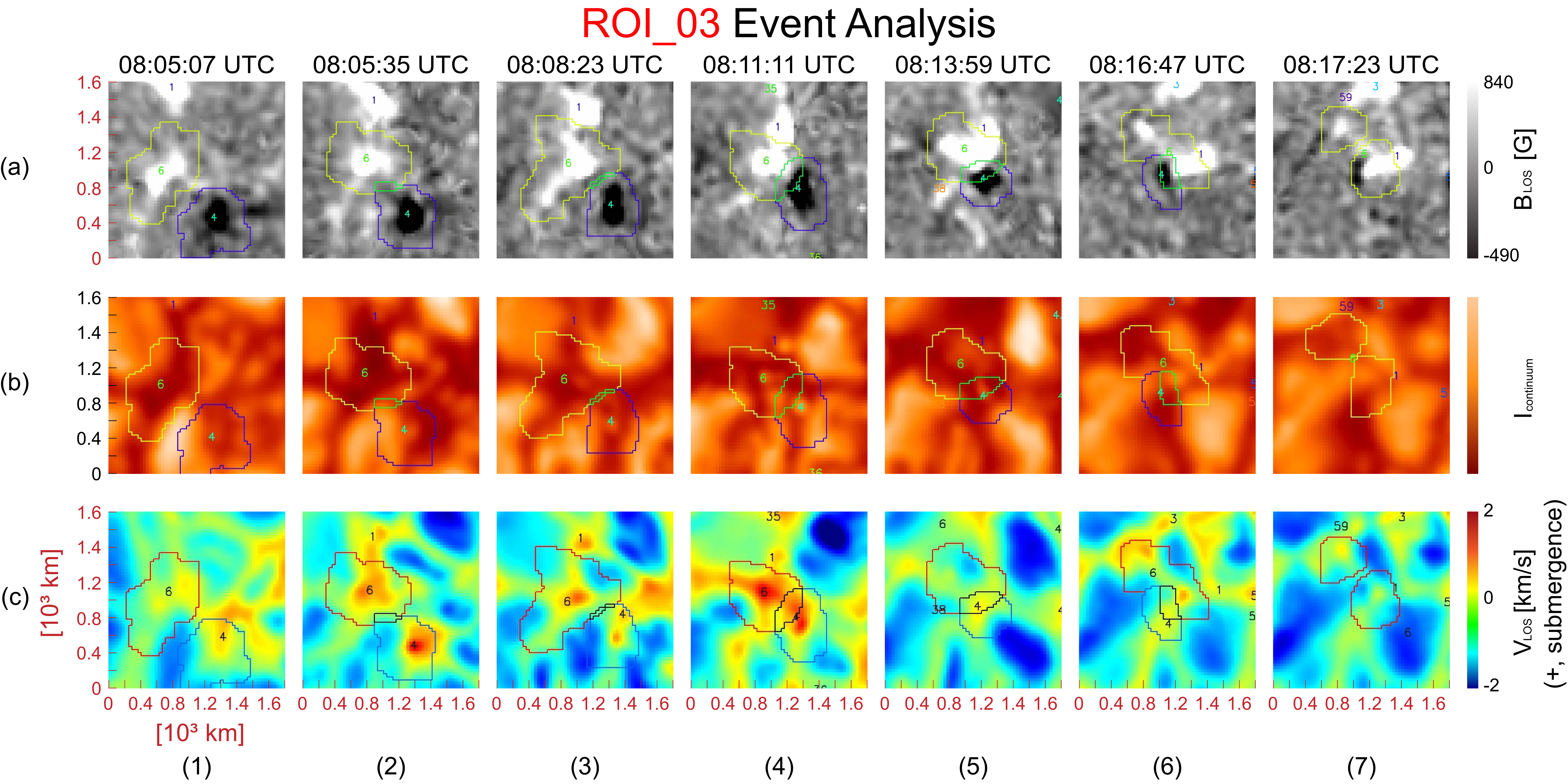}
\caption{Evolution of B$_\mathrm{LOS}$ (top series), 630.1 nm intensity (middle series) and $V_\mathrm{LOS}$ (bottom series). ROI$_{03}$ in the time series begins with panel 1 (time immediately before PIL is defined) and progresses to panel 7 (time immediately after PIL is no longer defined).
\label{fig:roi03fulldataset}}
\end{figure*}

Figure~\ref{fig:roi03fulldataset} (top row) shows evolution of the magnetic field in one example event, region of interest $3$ (ROI\_03). We chose to give this event special attention since it showed a marked flux cancellation, allowing us to compare the Doppler velocity in the PIL region before and during the cancellation. Furthermore, the polarities in this region were easy to identify by eye and there was very noticeable flux cancellation in both regions.


\begin{figure}[ht!]
\includegraphics[width=1.0\linewidth]{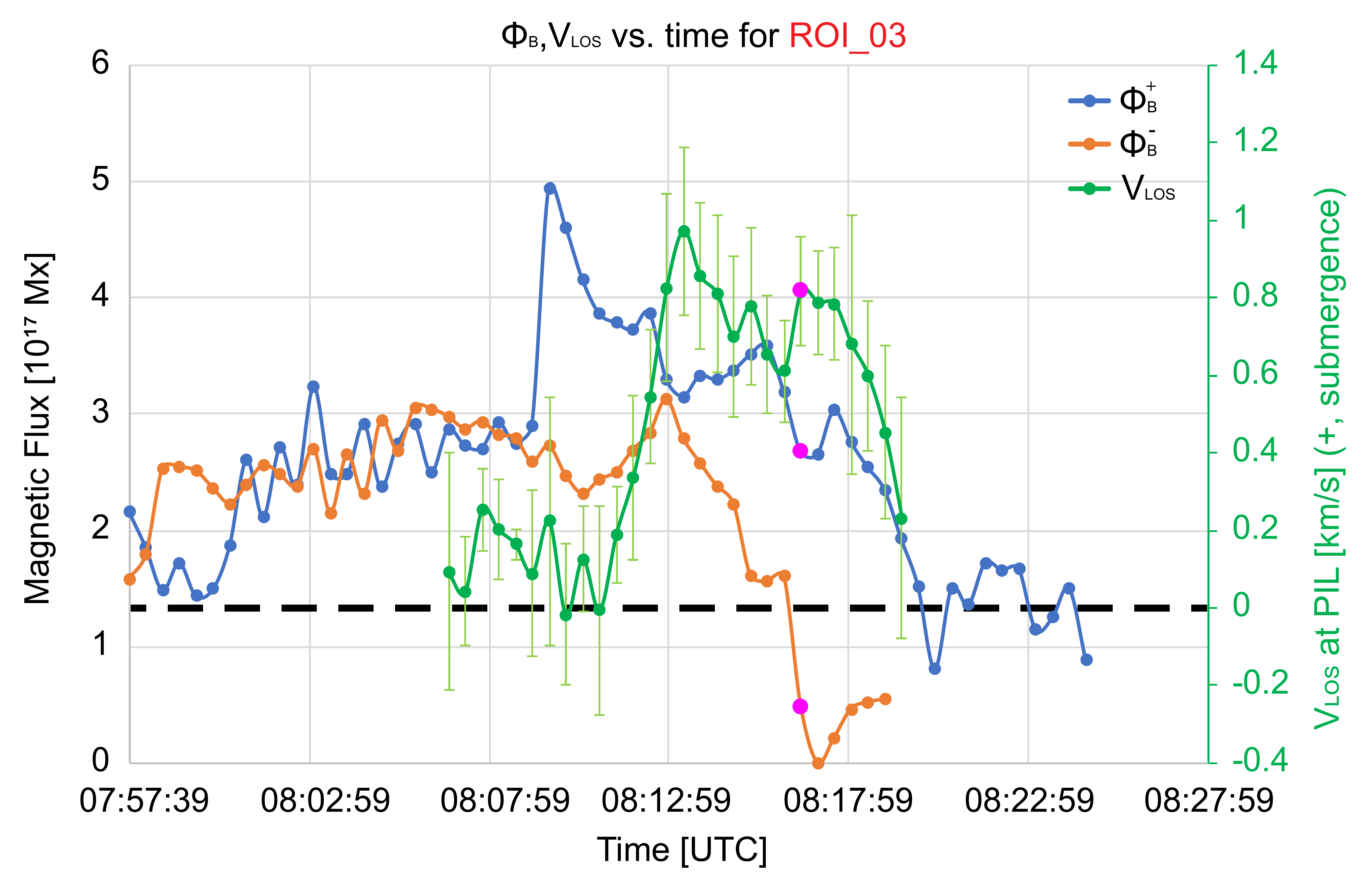}
\caption{Evolution of LOS magnetic flux $\mathrm\Phi_B$ and PIL mean Doppler velocity $V_\mathrm{LOS}$ in ROI\_03. Blue data points are $\Phi^{+}_{B}$; orange points are $\Phi^{-}_{B}$. The pink dots represent the time when maximum flux cancellation was achieved. The green data points are $V_\mathrm{LOS}$, and are only counted while the PIL is defined. Positive values of $V_\mathrm{LOS}$ represent submergence. A dotted line is placed at $V_\mathrm{LOS}$ = 0 km~s$^\mathrm{-1}$ to aid in viewing the diagram.
\label{fig:roi03plot}}
\end{figure}


From the beginning of its detection, ROI\_03 contained positive and negative polarities with relatively equal magnetic flux. Around 10 minutes into the tracking run, the two polarities became entangled (see PIL region in Figure~\ref{fig:roi03fulldataset} panel 2a). Around six minutes after the PIL is defined, or at around 08:11:11 UTC, the polarities began cancelling, losing flux at an average rate of 4.9$\mathrm\times$10$^{14}$ Mx s$^\mathrm{-1}$. $V_\mathrm{LOS}$ in the PIL region shows a small submergence of ~0.1 km~s$^{-1}$ when the PIL was first defined (Figure~\ref{fig:roi03fulldataset}, panel 2c) but then increased to over 1 km~s$^{-1}$ after the cancellation intensifies (Figure~\ref{fig:roi03fulldataset}, panel 4c). The event lasted ten minutes before the negative polarity decreased below the instrument noise level as seen in Figure~\ref{fig:roi03fulldataset}, panel 7a. During the cancellation the positive and negative polarities lost $1.87\times10^{17}$ Mx and $\mathrm2.48\times10^{17}$ Mx, respectively, i.e. around 54.7\% of the unsigned initial magnetic flux of the bipole. The cancellation took place in an intergranular lane as seen in the continuum and Doppler velocity images. As illustrated in Figure~\ref{fig:roi03plot}, before cancellation, the Doppler velocity of the PIL was $\mathrm \approx0.1$ km~s$^{-1}$. As both polarities started cancelling, with the peak cancellation rate of 4.4$\times 10^{15}$ Mx s$^{-1}$ occurring at 08:16:35 UTC, the downlflows at the PIL increased up to $\mathrm\approx$1 km~s$^{-1}$, persisting for approximately five minutes.

\subsection{Results: Statistical Analysis of 38 Cancellation Events}\label{sec:res_stat}

We repeat the analysis described in the previous section for $38$ events in our dataset. In Table~\ref{tab:summarytable} we present a summary of all events and major indices. In Table~\ref{tab:all38cancellations} we summarize our results showing cancellation parameters for $38$ cancellations events. 

{\it Duration:} We found an average cancellation duration of $\approx$ 39.2 minutes. Cancellation lifetimes of $9$ minutes were reported by \citet{Internetwork_3_Support_5} and \citet{CancellationLifeTimeGosic} with a range of 1 to 22 minutes found by \citet{Chromosphere_1}.

{\it Initial magnetic flux:} The distribution of initial fluxes, $\phi_{B,i}$, of the features is seen in Table~\ref{tab:all38cancellations}.
and we see a distribution clustered around $\mathrm[1-3]\times10^{17}$ Mx. The observed magnetic flux is $\mathrm[0.7-8.2]\times10^{17}$ Mx. This wide range of values shows that our dataset reflects the diversity of internetwork magnetic field strengths. The average value of ${\bar \Phi_{B,i}}=3.3\pm 0.3\times10^{17}$ Mx, and is lower than the value of $\mathrm5.5\times10^{17}$ Mx reported by \citet{Internetwork_3} and $6\times10^{17}$ Mx reported by \citet{Cancellation_3} but is nearly identical to the $3\times10^{17}$ Mx reported by \citet{InitialFluxMeasurementGosic}.

{\it Cancelled magnetic flux:} Values of the magnetic flux change, $\mathrm \Delta \Phi_{B}$, show a clustering below $\mathrm10^{17}$ Mx ranging from $\mathrm0.2\times10^{17}$ Mx to $3.1\times10^{17}$ Mx. We find the average cancelled magnetic flux to be $\mathrm \Delta \Phi _{B}=[1.6\pm 0.16]\times10^{17}$ Mx. Comparing with the initial flux, this change corresponds to $\mathrm[44.9\pm 2.3]\%$ of the initial flux being cancelled.

{\it Cancellation rates:} We found cancellation rates ranging from $\mathrm0.3\times10^{14}$ Mx s$^{-1}$ to $7.4\times10^{14}$ Mx s$^{-1}$ with a mean cancellation rate of $3.8\pm 0.5\times10^{14}$ Mx s$^{-1}$. \citet{Cancellation_1} found similar results for small magnetic elements with fluxes around 10$^{17}$ Mx and the flux decay rates of $4\times10^{14}$ Mx s$^{-1}$. Our values were slightly larger than the $\mathrm2.6\times10^{14}$ Mx s$^\mathrm{-1}$ found by \citet{Cancellation_3} and the $\mathrm3.6\times10^{14}$ Mx s$^\mathrm{-1}$ reported by \citet{CancellationRates}. For all events we found the mean of the peak flux cancellation rate to be $R_\mathrm{peak}=2.9\pm 0.2\times10^{15}$ Mx s$^\mathrm{-1}$, i.e. around one order of magnitude higher than the average rate, $\mathrm{R_{avg}}$.

{\it Cancellation rates averaged over PIL length:} We find the mean specific cancellation rate for all events to be $\mathrm{r= 2.7\pm 0.5\times10^{6}}$ G cm s$^{-1}$. Compared to \citet{Cancellation_1} (7.3$\mathrm \times10^{6}$ G cm s$^{-1}$) and \citet{Cancellation_4} (8$\mathrm \times10^{6}$ G cm s$^{-1}$), our results are around two times smaller, but greater than the values obtained by \citet{CancellationRates} (1.1$\mathrm \times10^{6}$ G cm s$^\mathrm{-1}$) and nearly identical to the results of \citet{SpecificCancellationRate} (2.32$\mathrm \times10^{6}$ G cm s$^{-1}$).

\begin{figure}[ht!]
\includegraphics[width=1.1\linewidth]{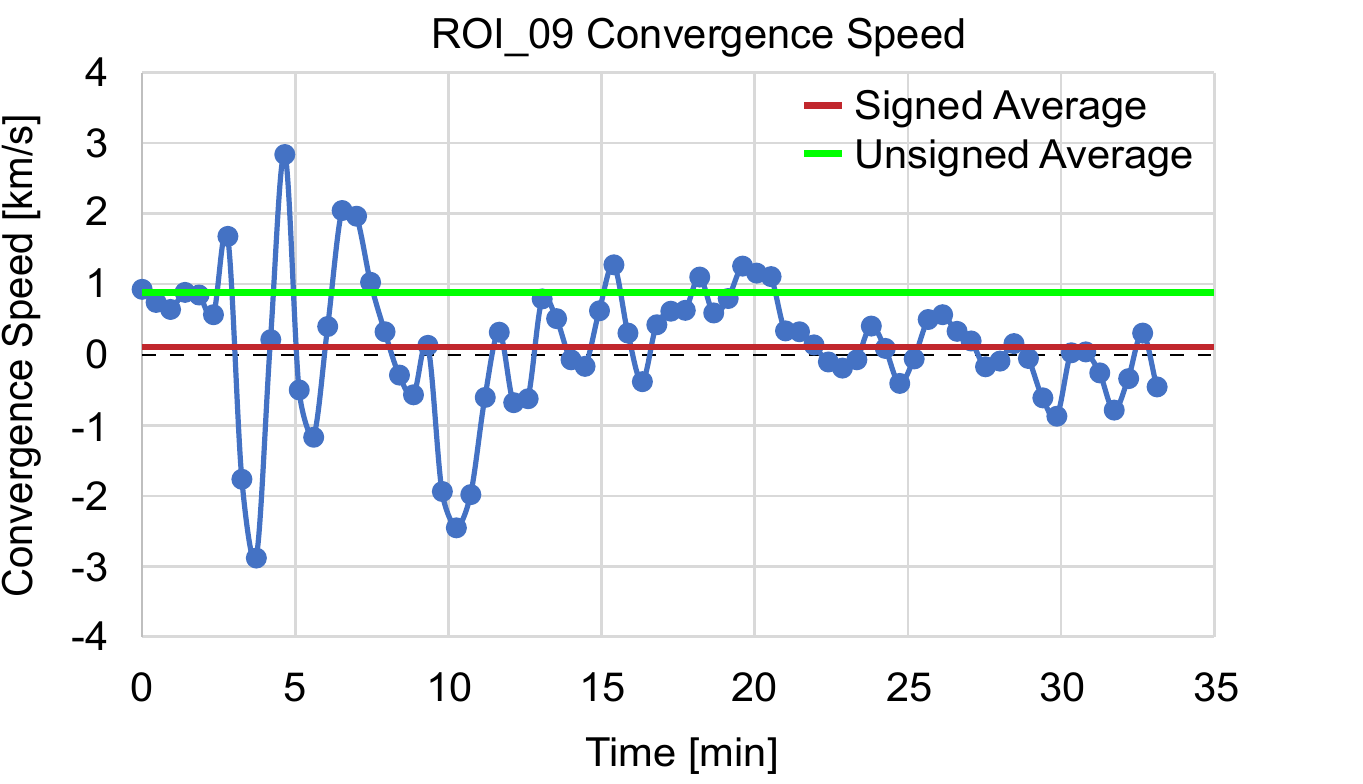}
\caption{Convergence speed, $v_\mathrm{conv}$, over the cancellation duration as calculated with the COG method for ROI\_09. As the cancelling features occasionally drift apart, the $v_\mathrm{conv}$ becomes negative. The average $v_\mathrm{conv}$ for that event, 0.1 km~s$^{-1}$ is represented by the red line. The proper motion speeds are calculated by $v\mathrm{v_{proper}}$= $\left |v_\mathrm{conv} \right |$, and the mean value is shown by the green line, which is 0.7 km~s$^{-1}$. ROI\_09 is shown here since it exhibits strong polarity drift.\label{fig:convspeed}}
\end{figure}

{\it Convergence speeds:} For $38$ cancellation events we find mean convergence speed $v_\mathrm{conv}=0.6\pm 0.07~km~s^{-1}$ with values ranging from $\mathrm0.2~km~s^{-1}$ to 2.1~km~s$^\mathrm{-1}$. \citet{Cancellation_1} reported a range of 0.3~km~s$^\mathrm{-1}$ to 1.8~km~s$^\mathrm{-1}$. These convergence speeds are similar to supergranular flow velocities found by \citet{SpecificCancellationRate,CancellationRates,supergranularmotion}. Still, we agree with \citet{Cancellation_1} that the wide distribution of convergence speeds as seen in individual cancellation events (see Figure~\ref{fig:convspeed}) are evidence of the more chaotic flows found on granular-scales. Using the average convergence speed to deduce whether cancellations are driven by supergranular motions or convective behavior is not entirely appropriate. According to Eq.~\ref{eqvel}, polarities moving away would have a negative convergence speed. Averaging the unsigned speeds ($v_\mathrm{proper}=|v_\mathrm{conv}|$) produces a result that more accurately depicts the internetwork environment. Using an approach similar to \citet{Cancellation_1} we derived the proper motion speeds of the cancellations to be $\mathrm0.9\pm0.05~km~s^{-1}$, which is markedly higher than the convergence speeds. These proper motion speeds are more consistent with the rms velocity of magnetic elements in internetwork areas found by \citet{rmsvelocity}. Our results are also consistent with other studies of magnetic cancellations, e.g. \citet{granularmotions} reporting proper motion speeds around $\mathrm1~km~s^{-1}$ while \citet{Cancellation_1} observed proper motion speeds higher than convergence speeds calculated using the traditional COG method (as seen in Figure~\ref{fig:convspeed}).
\citet{keys2011} also noted that the horizontal velocity of merging bright points is $\approx$ 1 km~s$^{-1}$ and higher than the speed of isolated bright points.

\begin{figure}[ht!]
\includegraphics[width=1.1\linewidth]{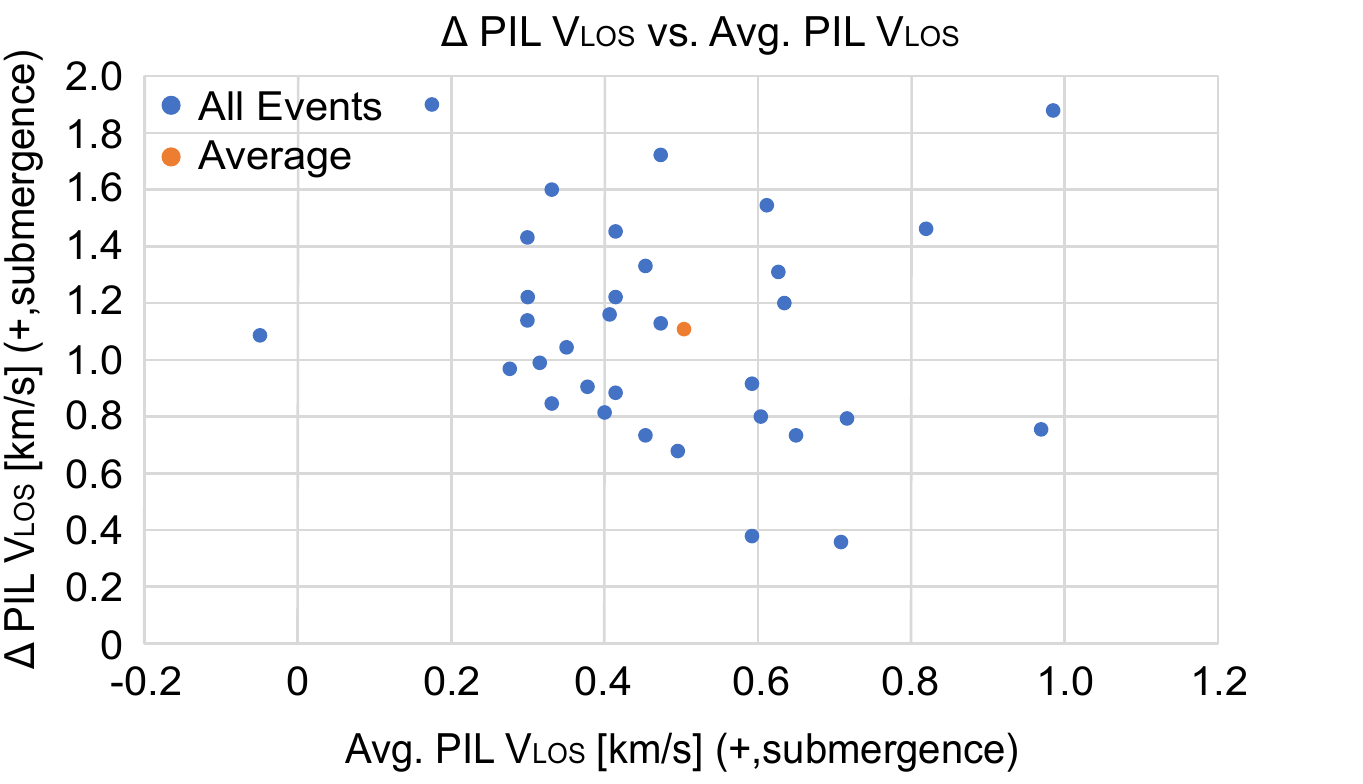}
\caption{Scatter plot of $\Delta$ $V_\mathrm{LOS}$ vs. $V_\mathrm{LOS,avg}$ for all events. Recall that $\Delta$ $V_\mathrm{LOS}$ is the difference between the extreme values of $V_\mathrm{LOS}$ over the lifetime of the event. \label{fig:deltavsavgvlos}}
\end{figure}

{\it Downflow speeds:} Across $38$ cancellations we observed average downflows of $V_\mathrm{LOS,avg}$ = 0.5$\pm 0.03$ km s$^{-1}$ and mean peak downflow speeds of $V_\mathrm{LOS,peak}$ = 0.6$\pm 0.1~km~s^{-1}$. These high downflow speeds observed at flux cancellation sites are consistent with results from \citet{downflows_omegaloop}. \citet{fluxsubmergence} found that emission structures from IN magnetic cancellations lasted longer in the photosphere than the chromosphere and corona, concluding that magnetic flux is retracting below the surface. Analyzing the magnetic elements overlaid on the continuum images (i.e. Figure~\ref{fig:roi03fulldataset} panels 1-7b), we noticed that many magnetic bipoles began cancellation in upflow regions and ended in downflow lanes, most likely due to convective motions. This was also observed in \citet{Cancellation_1}. To explore this phenomenon we compared the $\Delta V_\mathrm{LOS}$ to the $V_\mathrm{LOS,avg}$, shown in Figure~\ref{fig:deltavsavgvlos}. This would indicate whether small outlier values of $V_\mathrm{LOS,avg}$ simply resulted from a greater downflow shift from local upflow convection. We see that although the mean $V_\mathrm{LOS,avg} = 0.5\pm 0.03$, the $\Delta V_\mathrm{LOS}= 1.1 \pm 0.07$. There is no statistically significant relationship between $\Delta V_\mathrm{LOS}$ and $V_\mathrm{LOS,avg}$ (The Pearson Correlation Coefficient, $\rho=0.07$).

\begin{figure}[ht!]
\includegraphics[width=1.1\linewidth]{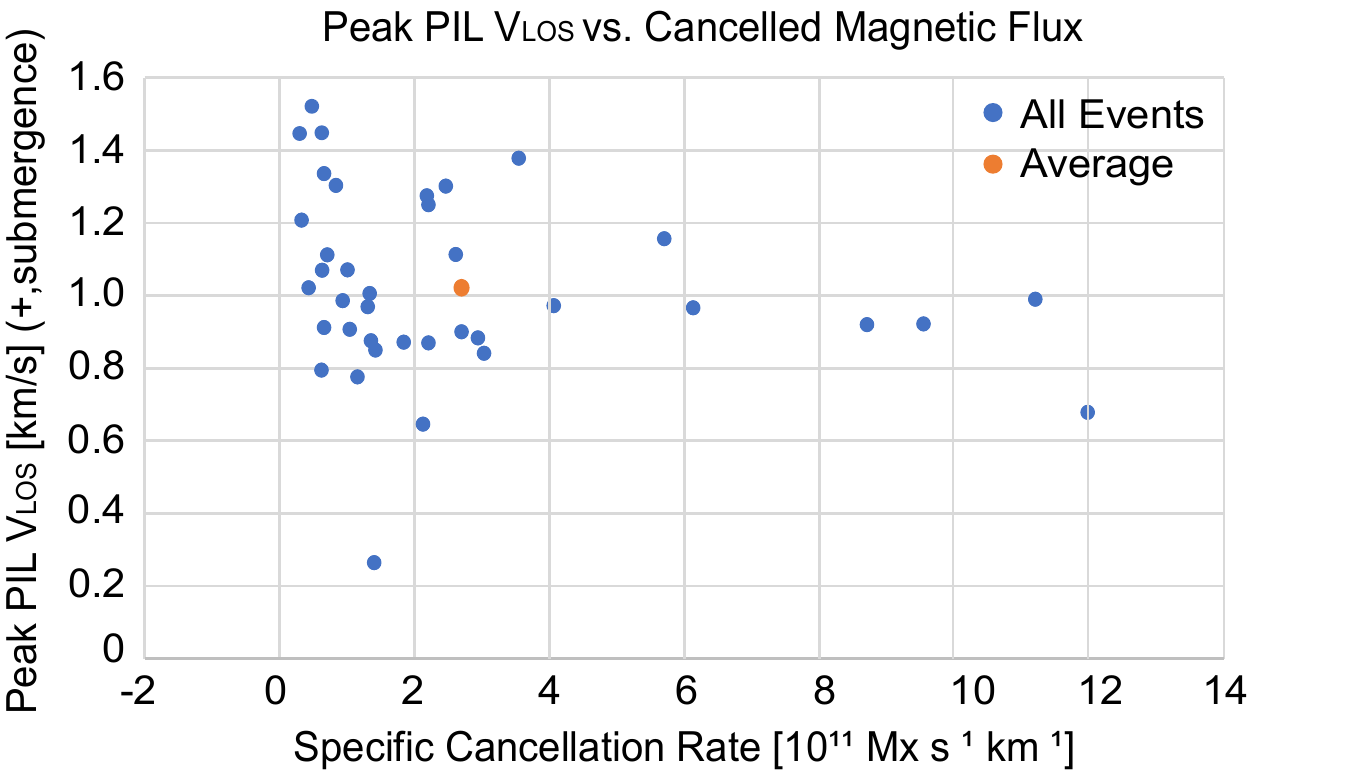}
\caption{Scatter plot of peak Doppler velocity, $V_\mathrm{LOS,peak}$ vs. total cancelled flux, $\Delta \Phi_{B}$ for all events. \label{fig:NEWFIGGG}}
\end{figure}

Investigating further correlations between $V_\mathrm{LOS}$ and other statistical parameters, we first addressed the hypothesis that more magnetic flux cancellation may lead to a higher downflow signature. We plotted $V_\mathrm{LOS,peak}$ vs. $\mathrm \Delta \Phi _{B}$ in Figure~\ref{fig:NEWFIGGG} and no correlation was found. Relationships between $\Delta$ $V_\mathrm{LOS}$ and $\mathrm \Delta \Phi _{B}$, and $V_\mathrm{LOS,avg}$ vs. $\mathrm \Delta \Phi _{B}$ were also plotted but are not shown in the manuscript; analysis found no correlation. R$^{2}$ values were 0.04, 0.004, and 0.03 for the data sets, respectively. The notion that downflows may be limited by cancelled flux, as if by a threshold, is not supported by these data. This finding may also suggest that the magnitude of downflows is more dependent on the magnitude of decay than its amount.

To investigate this notion, we plotted all $V_\mathrm{LOS}$ parameters against the average and peak flux cancellation rates, R$_\mathrm{avg}$ and R$_\mathrm{peak}$, respectively.
We find that while there is statistical evidence to support cancellation events drive submergence, whether or not the submergences are a result of the magnetic cancellation or the typical downflows found in intergranular lanes where many polarities eventually coalesce and cancel is indeterminable from this initial analysis. As opposed to studying the effects of $\Delta \Phi _{B}$ on $V_\mathrm{LOS}$, R$_\mathrm{avg}$ and R$_\mathrm{peak}$ provide the best insight into how the physical process of magnetic cancellation relates to submergences, since timescales vary with differing amounts of $\mathrm \Delta \Phi _{B}$. Compared to typical submergence speeds in intergranular lanes, which have been reported as 0.30 to 0.50 km~s$^\mathrm{-1}$ by \citet{intergranularflow}, the downflows associated with the cancellation events are much faster.

\begin{figure}[htb!]
\includegraphics[width=1.1\linewidth]{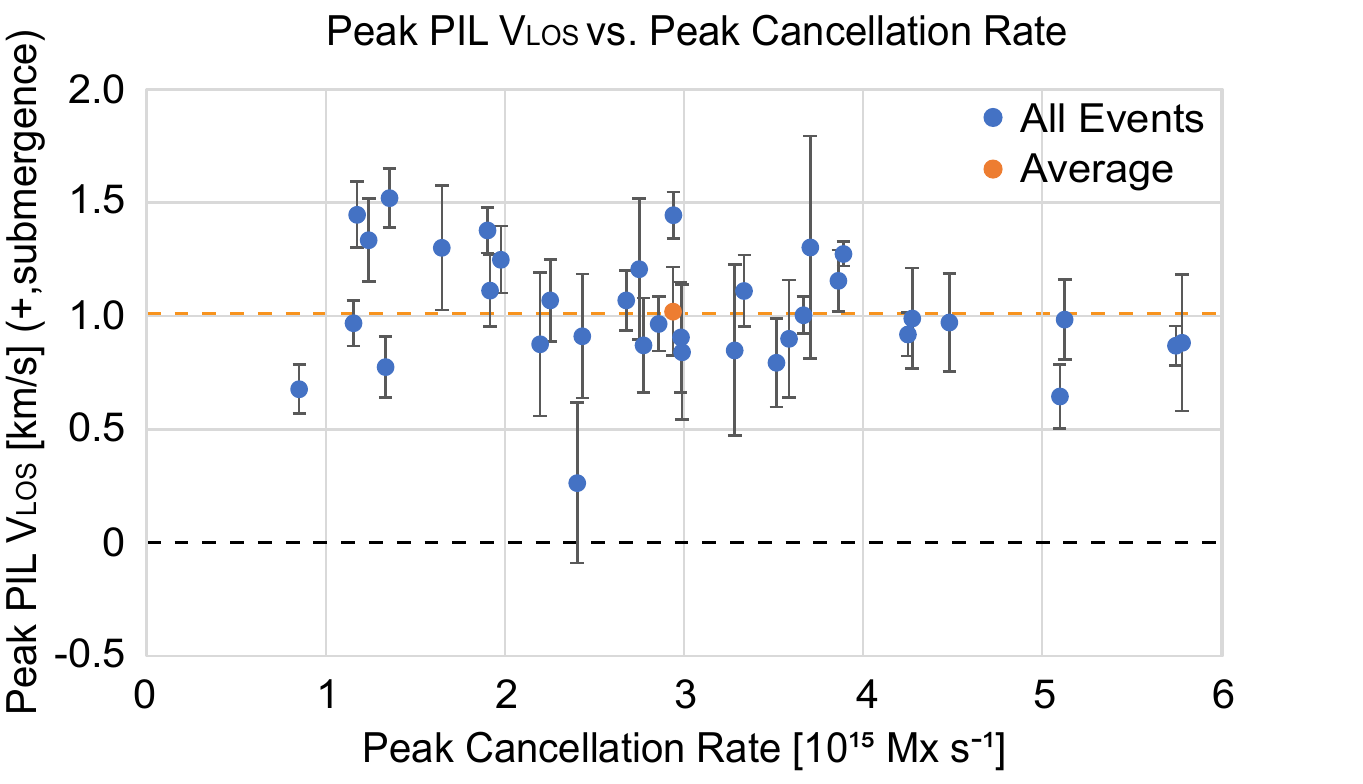}
\caption{Scatter plot of peak Doppler velocity, $V_\mathrm{LOS,peak}$ vs. peak flux cancellation rate, $R_\mathrm{peak}$ for all events. Recall that positive values of $R_\mathrm{peak}$ indicate flux cancellation while negative values indicate flux accumulation. \label{fig:peakvlosvsmaxrate}}
\end{figure}

\begin{figure}[htb!]
\includegraphics[width=1.0\linewidth]{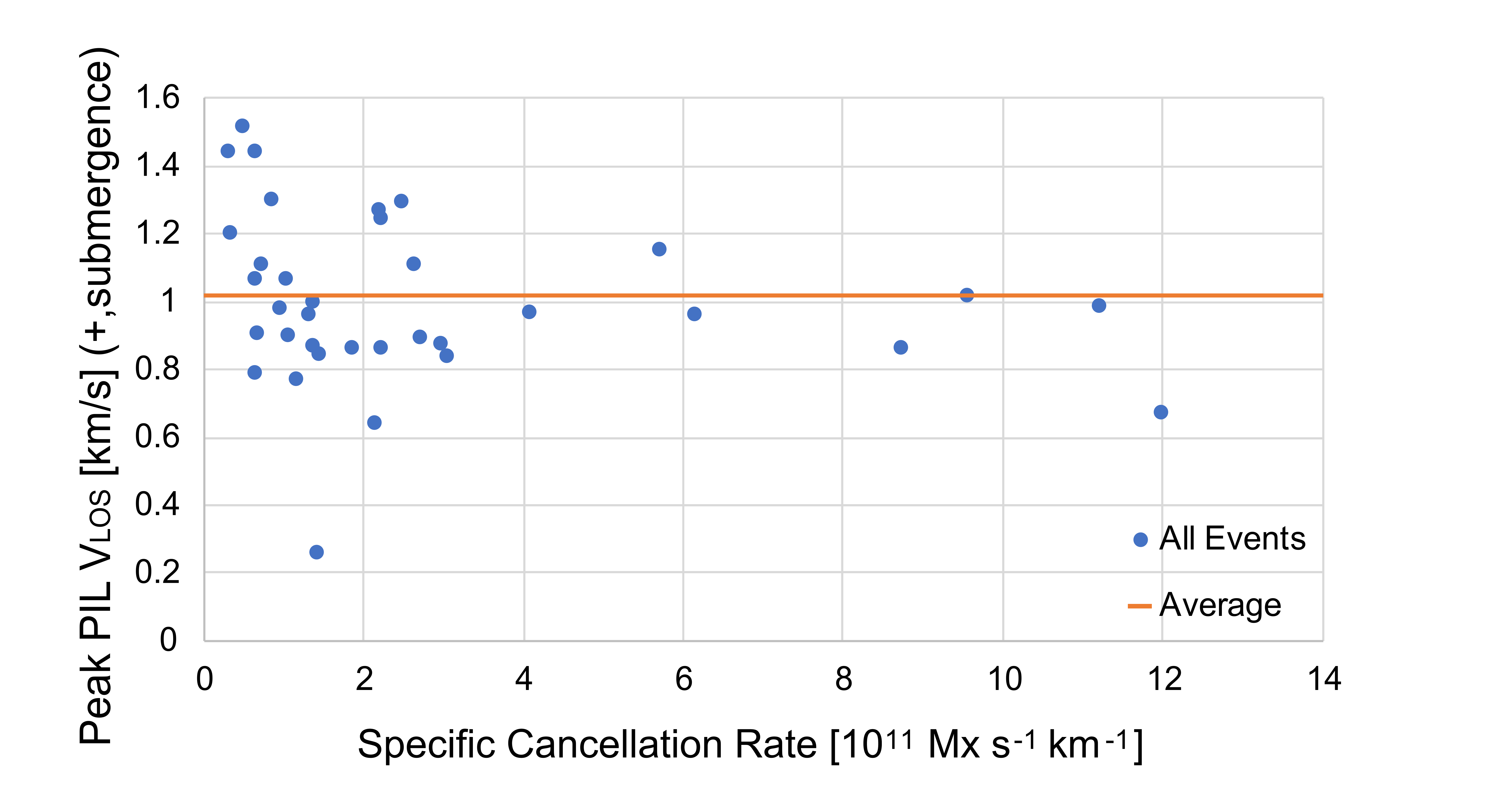}
\caption{Scatter plot of peak Doppler velocity, $V_\mathrm{LOS,peak}$ vs. specific cancellation rate, r for all events. Recall that r measures cancellation per. unit length of the PIL. \label{fig:peakvlosvsflux}}
\end{figure}

For $V_\mathrm{LOS}$ vs. r, the specific cancellation rate, we find no correlation (Figure~\ref{fig:peakvlosvsflux}).

\begin{figure}[ht!]
\includegraphics[width=1.1\linewidth]{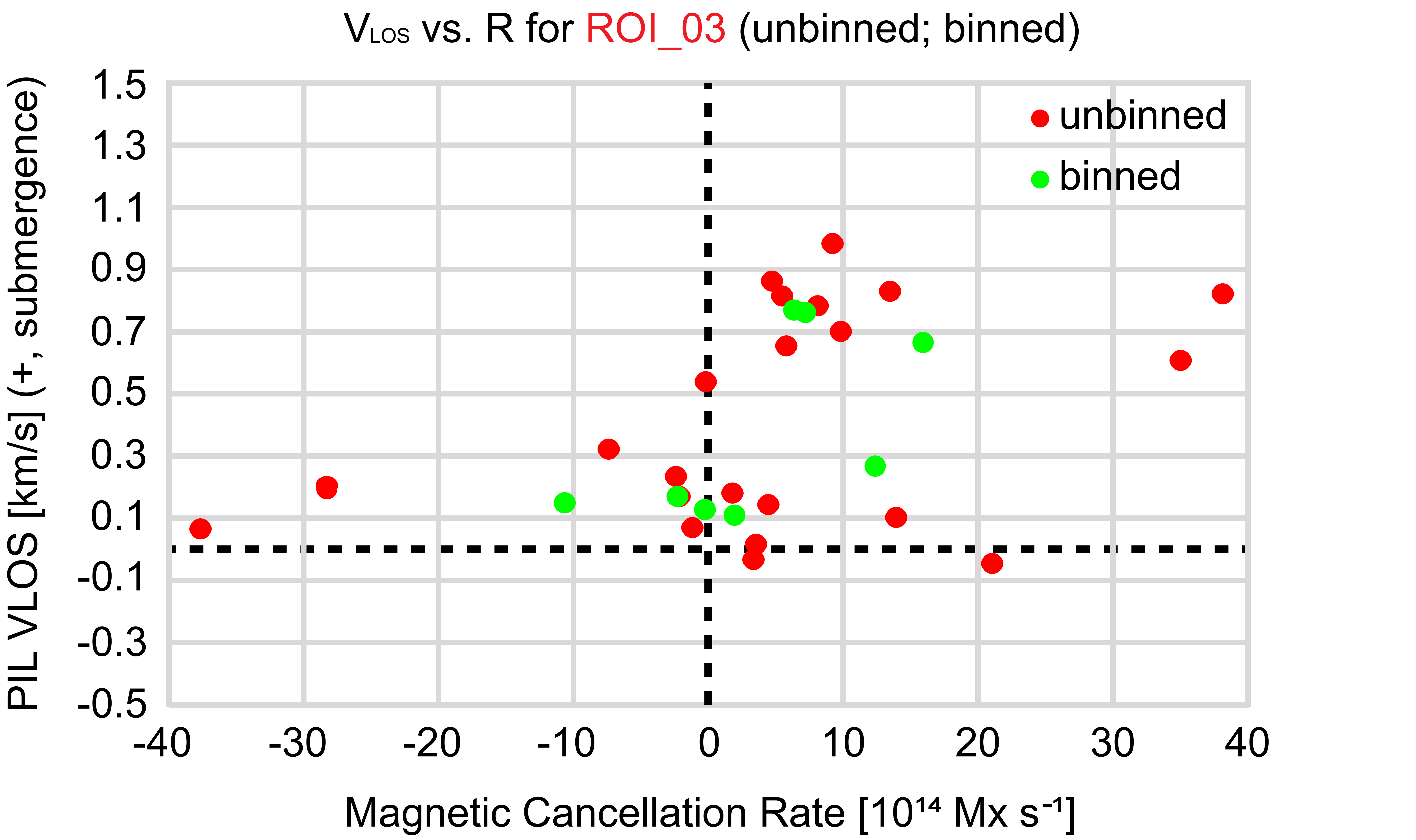}
\caption{Scatter plot of Doppler velocity, $V_\mathrm{LOS}$ vs. flux cancellation rate, R for ROI\_03. Unbinned and binned values are indicated. \label{fig:roi_03 binnedvsnot}}
\end{figure}

Analyzing single event-performance we found a stronger correlation between peak Doppler velocity and specific cancellation rate than analysis of the entire cancellation set. In ROI\_03 (see Figure~\ref{fig:roi03fulldataset}) we directly plotted R(t) vs. $V_\mathrm{LOS}$, finding weak correlation. We hypothesize that this might be caused by the definition of magnetic cancellation stating that only one feature in the cancelling pair is required to lose magnetic flux; the other participant may gain magnetic flux concurrently, therefore R, which we define as R(t)=$\mathrm R(t)^+$+$R(t)^-$, the net decay of the bipole, is not entirely appropriate for these cases. Secondly, the measurement variability is so great that its effect can directly confound results. Given these postulates, from the first time step where cancellation occurred we averaged every 
84 seconds, essentially time-binning the values of R and $V_\mathrm{LOS}$ by 3. We find a weak correlation of unbinned values ($\mathrm R^2=0.1558$) yet a moderate correlation with the binned values ($\mathrm R^2=0.3779$). This is seen more clearly in Figure~\ref{fig:roi_03 binnedvsnot}.

As stated earlier, during cancellation events we see magnetic fields frequently below the instrument noise level, so it is possible a correlation between $V_\mathrm{LOS}$ and $R$ would be more apparent with a more sensitive dataset. We also analyzed the correlation between $V_\mathrm{LOS}$ and $R$ for the 15, 10, and 5 strongest flux patches but found no correlation. It is possible that some undiscovered mechanism is creating a positive time separation between submergences and magnetic cancellation such that when the features dip below the instrument threshold even more of the downflows are lost due to background noise.

We calculated error by taking the standard deviation divided by the square root of the number of frames in the sequence relevant to the cancellation event. Ultimately, the standard error represents the standard deviation of the mean within the dataset.

Finally, we find that specific cancellation rate, r, was correlated with R$_\mathrm{avg}$ (R$^{2}$=0.707). This indicates that the primary transport mechanism of magnetic flux out of the bipole is through the polarity inversion line.

\begin{table*} \begin{center}
\begin{tabular}{cccc}
Variable & This work & \citet{Cancellation_1} & \citet{CancellationRates}\\
\hline
\hline
$\mathrm \Phi _{B,i}$ [$10^{17}$ Mx] & 3.3$\pm 0.3$ & $\approx$1.0 & 25\\
$\mathrm \Delta \Phi _{B}$ [$10^{17}$ Mx] & 1.6$\pm 0.2$ & & \\
$\mathrm \Delta \Phi _{B}$ [\%$^{a}$] & 44.9$\pm 2.3$ & 80 & \\
$R_{avg}$ [$10^{14}$ Mx s$^{-1}$] & 3.8$\pm 0.5$ & 4.0 & 3.6\\
$R_{peak}$ [$10^{15}$ Mx s$^{-1}$] & 2.9$\pm 0.2$ & & \\
r [10$^{6}$ G cm s$^{-1}$] & 2.7$\pm 0.5$ & 7.3 & 1.1\\
$\mathrm{T}$ [min] & 39.2 & 3.3 & \\
$V_\mathrm{LOS,avg}$ [km s$^{-1}$] & 0.5$\pm 0.03$ & & \\
$V_\mathrm{LOS,peak}$ [km s$^{-1}$] & 1.0$\pm 0.04$ & 1.4 & \\
$\mathrm \Delta V_{LOS}$ [km s$^{-1}$] & 1.1$\pm 0.03$ & & \\
v$_\mathrm{conv}$ [km s$^{-1}$] & 0.6$\pm 0.06$ & [0.3-1.8] & 0.35\\
\end{tabular}
\caption{Summary table showing mean values for 38 magnetic cancellations; $\mathrm \Phi _{B,i}$ is the initial flux of the bipole, $\mathrm \Delta \Phi _{B}$ is the amount of cancelled flux, $\mathrm R_{avg}$ is the average rate of cancellation, $\mathrm R_{peak}$ is the peak rate of cancellation, r is the specific cancellation rate, T is the duration of the event, $V_\mathrm{LOS,avg}$ is the average Doppler velocity at the PIL, $V_\mathrm{LOS,peak}$ is the peak Doppler velocity at the PIL, $\mathrm \Delta V_{LOS}$ is the total change in Doppler velocity at the PIL, v$_\mathrm{conv}$ is the convergence velocity of the polarities, and $^{a}$ represents the percent of initial flux cancelled; many cancellations ended below the noise floor.}
\normalsize \label{tab:summarytable} \end{center} \end{table*}

\begin{table*} \begin{center}
\begin{tabular}{cccccccccc}
ROI & $\mathrm \Phi _{B,i}$ & $\mathrm \Delta \Phi _{B}$ & $\mathrm R_{avg}$ & $\mathrm R_{peak}$ & r & T & $V_\mathrm{LOS,peak}$ & $\mathrm \Delta V_{LOS}$ & v$_\mathrm{conv}$\\
$\#$ & [$10^{17}$ Mx] & [$10^{17}$ Mx] & [$10^{14}$ Mx s$^{-1}$] & [$10^{15}$ Mx s$^{-1}$] & [10$^{6}$ G cm s$^{-1}$] & $[$min$]$ & [km s$^{-1}$] & [km s$^{-1}$] & [km s$^{-1}$]\\
\hline
1 & 5.6 & 3.0 & 5.4 & 2.7 & 1.8 & 9.8 & 0.8 & 0.8 & 0.4 \\
2 & 1.9 & 0.7 & 0.9 & 2.7 & 0.3 & 8.4 & 1.2 & 1.1 & 0.1 \\
3 & 5.8 & 2.6 & 4.5 & 4.4 & 4.0 & 10.3 & 0.9 & 0.9 & 0.4 \\
4 & 3.1 & 3.4 & 1.8 & 3.7 & 0.8 & 23.8 & 1.3 & 1.3 & 0.2 \\
5 & 6.9 & 1.7 & 13.0 & 4.2 & 8.7 & 9.3 & 0.9 & 1.6 & 0.4 \\
6 & 5.9 & 2.0 & 8.1 & 2.9 & 3.0 & 5.1 & 0.8 & 0.3 & 0.9 \\
7 & 5.2 & 2.5 & 3.7 & 2.9 & 1.0 & 13.5 & 0.9 & 1.1 & 0.2 \\
8 & 6.1 & 3.0 & 2.5 & 5.1 & 0.9 & 19.1 & 0.9 & 1.1 & 0.4 \\
9 & 2.5 & 1.2 & 0.6 & 2.9 & 0.2 & 33.6 & 1.4 & 2.0 & 0.1 \\
10 & 6.9 & 2.9 & 4.6 & 5.7 & 2.2 & 9.3 & 0.8 & 1.5 & 0.4 \\
11 & 8.2 & 2.5 & 3.5 & 3.2 & 1.1 & 10.3 & 0.8 & 0.8 & 0.3 \\
12 & 5.5 & 2.6 & 5.9 & 4.2 & 11.2 & 9.3 & 0.9 & 0.7 & 0.5 \\
13 & 1.3 & 0.5 & 1.3 & 3.3 & 0.7 & 6.7 & 1.1 & 0.9 & 0.6 \\
14 & 2.1 & 1.4 & 1.9 & 2.2 & 1.0 & 10.3 & 1.0 & 1.4 & 0.3 \\
15 & 3.3 & 1.8 & 3.9 & 3.8 & 2.1 & 11.2 & 1.2 & 1.5 & 0.6 \\
16 & 1.9 & 1.0 & 6.0 & 1.6 & 2.4 & 3.7 & 1.3 & 0.7 & 1.3 \\
17 & 2.0 & 0.6 & 4.3 & 1.9 & 2.2 & 3.7 & 1.2 & 1.9 & 1.0 \\
18 & 2.6 & 1.6 & 3.0 & 3.6 & 1.3 & 10.3 & 1.0 & 0.9 & 0.4 \\
19 & 5.2 & 3.0 & 7.2 & 5.7 & 2.9 & 6.1 & 0.8 & 0.7 & 0.7 \\
20 & 2.3 & 1.1 & 5.1 & 3.6 & 5.7 & 6.5 & 1.1 & 1.7 & 0.9 \\
21 & 1.0 & 0.3 & 6.0 & 8.5 & 11.2 & 3.7 & 0.6 & 0.8 & 0.7 \\
22 & 1.0 & 0.2 & 0.9 & 1.3 & 0.4 & 6.5 & 1.5 & 1.4 & 0.5 \\
23 & 1.4 & 0.6 & 0.9 & 2.6 & 0.6 & 11.7 & 1.0 & 1.4 & 0.3 \\
24 & 2.9 & 1.3 & 2.6 & 1.9 & 2.6 & 4.7 & 1.1 & 0.7 & 1.2 \\
25 & 5.7 & 2.9 & 3.9 & 3.5 & 2.7 & 15.4 & 0.9 & 1.2 & 0.1 \\
26 & 4.0 & 1.7 & 3.2 & 2.8 & 6.1 & 10.7 & 0.9 & 1.2 & 0.6 \\
27 & 9.4 & 0.5 & 2.2 & 1.2 & 3.7 & 5.1 & 1.3 & 1.3 & 0.8 \\
28 & 3.2 & 0.6 & 1.8 & 2.4 & 0.6 & 10.7 & 0.9 & 1.0 & 0.4 \\
29 & 4.1 & 1.4 & 3.3 & 2.2 & 1.3 & 7.0 & 0.8 & 0.6 & 0.7 \\
30 & 1.2 & 0.7 & 1.0 & 1.1 & 0.6 & 7.0 & 1.4 & 1.3 & 0.6 \\
31 & 7.5 & 0.3 & 2.7 & 1.3 & 1.2 & 3.2 & 0.7 & 0.3 & 1.1 \\
32 & 1.2 & 0.5 & 1.3 & 1.1 & 1.3 & 7.0 & 0.9 & 0.8 & 0.7 \\
33 & 2.8 & 1.3 & 2.9 & 2.4 & 1.4 & 5.6 & 0.2 & 1.0 & 0.7 \\
34 & 2.3 & 0.7 & 9.4 & 3.5 & 0.6 & 11.2 & 0.7 & 1.1 & 1.0 \\
35 & 3.5 & 2.3 & 4.2 & 5.1 & 2.1 & 10.3 & 0.6 & 0.9 & 0.3 \\
36 & 3.4 & 2.3 & 1.4 & 3.8 & 9.5 & 2.8 & 0.9 & 0.3 & 2.1 \\
37 & 1.9 & 0.1 & 2.0 & 1.9 & 0.4 & 3.2 & 1.0 & 0.5 & 0.8 \\
38 & 4.9 & 0.7 & 3.8 & 3.8 & 3.5 & 3.2 & 1.3 & 1.5 & 0.2 \\
\end{tabular}
\caption{Key parameters for all 38 magnetic cancellations.}
\normalsize \label{tab:all38cancellations} \end{center} \end{table*}

\section{Discussion and Conclusion}
\label{sec:dis&conc}
In this study we used spectropolarimetric measurements from the SST to investigate the physical properties of magnetic cancellations in the quiet Sun photosphere. Examining an LOS magnetogram we visually identified 38 cancellations, and using the YAFTA program suite \citep{YAFTA_1}, we tracked the magnetic elements involved in the cancellation and extracted their time-dependent physical parameters. We found that cancellations and downflows occur simultaneously, with an average relative speed of $\mathrm \Delta V_{LOS}$ = 1.1 km~s$^{-1}$. We found no increase of the linear polarization signal in our data, corresponding to the horizontal component of magnetic field B$_{t}$ probably because it was below the noise level of the observations. This means that we can not comment on whether the flux retracts below the photosphere and forms structures which could lead to reconnection. Our findings are consistent with the study from \citet{Kubo} which also did not find horizontal magnetic fields.
A snapshot of the data is included in Sec.~\ref{sec:appendix}. We expect that with more sensitive data these magnetic cancellations would be observed with enhanced $\mathrm{B_{t}}$ signals, providing stronger evidence of possible magnetic reconnection.

While establishing the link between downflows and cancellations is an important discovery in this study, we also calculated other statistical parameters that characterize these events: $\Phi _{B,i}$, $\Delta \Phi _{B}$, $R_{avg}$, $R_{peak}$, r, $\mathrm{T}$, and v$_\mathrm{conv}$ (see Table ~\ref{tab:summarytable}).

Aside from v$_\mathrm{conv}$, we estimated the proper motion of the magnetic elements by using the unsigned average of their speeds, $v_{proper}$= $\left |v_\mathrm{conv} \right |$. While this is not the same method used by others such as \citet{Cancellation_1}, we found v$_{proper}$ to be on average 1 km~s$^{-1}$. This was significantly higher than the average speed of v$_\mathrm{conv}$ = 0.6$\pm 0.06$. Because the $v_\mathrm{conv}$ for individual events was highly variable (see Figure~\ref{fig:convspeed}) and our new value of v$_\mathrm{proper}$ agreed with the rms velocity for magnetic internetwork elements \citep{rmsvelocity}, we assume that magnetic cancellations are driven by granular motions that force the magnetic elements into intergranular lanes where they cancel. In future studies, we would like to analyze the contribution of supergranular flows to the movement of the magnetic elements. Our exact value of v$_{proper}$ is likely an underestimate since it is still a relative measurement.

Although our study identified many cancellations, their lifetimes were noise-limited. In Section~\ref{sec:dis&conc} we theorized that because of the instrument noise level some weak fields may not have been detected and tracked by the YAFTA program. This is evident by only around 45\% of initial flux being cancelled in an average magnetic bipole. Given the YAFTA parameters outlined in Section \ref{sec:meth}, we were unable to track polarities below the noise floor of the dataset \citep{lambetal}. Because many of the events ended below the noise floor, this limited our analysis into the direct correlation between magnetic flux cancellation and plasma downflows.

Limiting our study to same-sized opposite-polarity elements and avoiding elements where same-signed flux recombined prevented us from detecting more events and thus led to an underestimation of magnetic flux evolution. In Figure 2 of \citet{Internetwork_1} the authors present a methodology to track differently sized features. We may employ a similar method in future studies of IN magnetic elements.

Although we determined the optimal threshold to detect features in YAFTA, more in-depth understanding of QS magnetism would be achieved by observations with larger-aperture telescopes, and more advanced tracking algorithms. In the YAFTA program we empirically determined both a minimum size and magnetic threshold that determined whether features were tracked. We did this by simply observing the point at which noise patterns were no longer detected as features by YAFTA, then used that threshold in the analysis of all the ROIs. As stated before, we excluded pixels below 40 G and magnetic elements under 4 pixels in size from our analysis. It is possible that algorithmically determining the exact thresholds for each ROI would yield better detection of the events near the end of their lifetimes.

Magnetic reconnection also may occur with U-loop emergences which cause brightenings in the local chromosphere (\citet{Chromosphere_1}, \citet{Chromosphere_2}, and others). In order to investigate the effects of small-scale reconnection events on the higher layers of the atmosphere, we plan to combine observations in photospheric (as those employed in this study) and chromospheric lines. In particular, we plan to acquire spectropolarimetric observations in \ion{Ca}{2} 854.2 nm to investigate the evolution of the chromospheric magnetic field and broad-band images in \ion{Ca}{2} UV lines (e.g. the \ion{Ca}{2}-H filter at the SST) to estimate the amount of energy released in the chromosphere during reconnection.

The statistical parameters found in our study
are important for implicating magnetic cancellations in future studies of the quiet Sun and complement existing literature. Lastly, it would be interesting to re-examine QS magnetic fields using higher-aperture telescopes. Higher-aperture telescopes naturally have higher spatial resolution which is required to resolve the PIL; better spectropolarimetric sensitivity is necessary to track features for longer times, and will increase the detectability of features and therefore increase the quality of the derived statistics. Spectropolarimetric measurements from existing and upcoming installments such the Big Bear Solar Observatory, the European Solar Telescope (EST) and the Daniel K. Inouye Solar Telescope (DKIST, \citealt{Rast2021}), respectively, will allow for more detailed study of the evolution of the magnetic field components.

\subsubsection*{Acknowledgements}
This work was supported by NSF through award number 1659878 for the Boulder Solar Alliance REU in 2019. FZ acknowledges support from the European Union’s Horizon 2020 research and innovation programme under the grant agreements no. 739500 (PRE-EST project) and no. 824135 (SOLARNET project), by the Universita` degli Studi di Catania (PIA.CE.RI. 2020-2022 Linea 2), by the Italian MIUR-PRIN grant 2017APKP7T on “Circumterrestrial Environment: Impact of Sun-Earth Interaction”. The National Solar Observatory is operated by the Association of Universities for Research in Astronomy, Inc. (AURA) under cooperative agreement with the National Science Foundation.
The Swedish 1-m Solar Telescope is operated on the island of La Palma by the Institute for Solar Physics of Stockholm University in the Spanish Observatorio del Roque de los Muchachos of the Instituto de Astrof{\'\i}sica de Canarias.
The Institute for Solar Physics is supported by a grant for research infrastructures of national importance from the Swedish Research Council (registration number 2017-00625).
This research is supported by the Research Council of Norway, project number 250810, and through its Centres of Excellence scheme, project number 262622.
We made much use of NASA's Astrophysics Data System Bibliographic Services. We wish to thank the US Taxpayers for their generous support for this project.

\appendix

\section{Appendix} \label{sec:appendix}
In this section we describe the evolution of four more cancellations in detail - ROI\_07, ROI\_08, ROI\_11, and ROI\_23. These events are shown in context with the dataset in Figures \ref{fig:observations} and \ref{fig:yaftasequence}.

\subsection{Analysis of ROI\_07}

\begin{figure*}[ht!]
\includegraphics[width=1.0\linewidth]{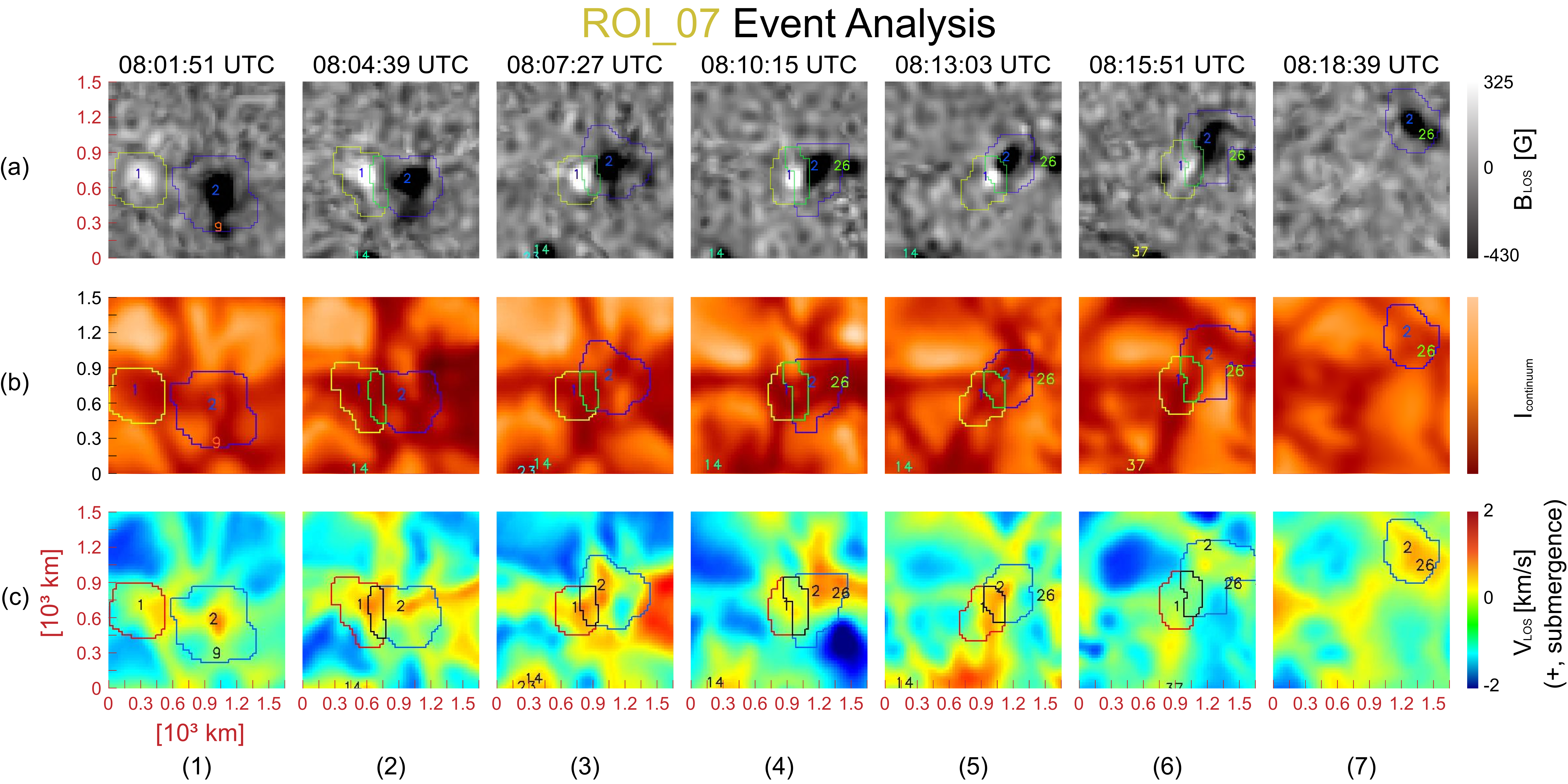}
\caption{Evolution of B$_\mathrm{LOS}$ (top series), 630.1 nm intensity (middle series) and $V_\mathrm{LOS}$ (bottom series). ROI\_{07} in the time series begins with panel 1 (time immediately before PIL is defined) and progresses to panel 7 (time immediately after PIL is no longer defined).
\label{fig:roi07fulldataset}}
\end{figure*}

\begin{figure}[ht!]
\centering
\includegraphics[width=.5\linewidth]{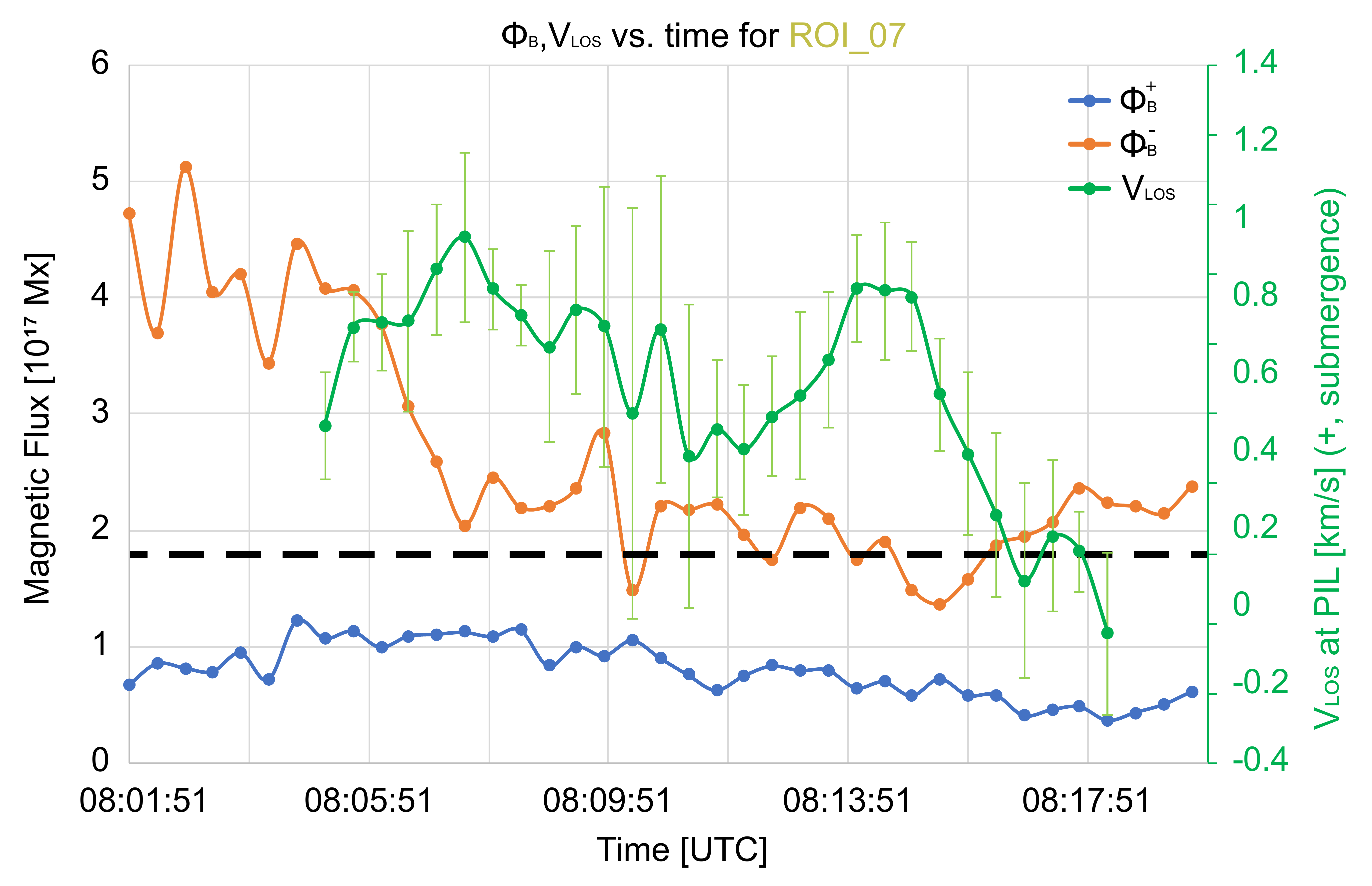}
\caption{Evolution of LOS magnetic flux and PIL mean Doppler velocity, $\Phi_\mathrm{B}$ and $V_\mathrm{LOS}$, in ROI\_07. Refer to the caption of Figure~\ref{fig:roi03plot} for an explanation of the legend and other graphical details.
\label{fig:roi07plot}}
\end{figure}

ROI\_07 involves a larger negative polarity and smaller positive polarity cancelling in an intergranular lane, as illustrated in Figure~\ref{fig:roi07fulldataset}. $\mathrm \Phi_{B,i}^{-}$ was $4.1\times10^{17}$ Mx and $\mathrm \Phi_{B,i}^{+}$ was $1.1\times10^{17}$ Mx. As shown in Figure~\ref{fig:roi07plot}, after approximately 13 minutes the bipole lost 49.7\% of its initial magnetic flux, or 2.6$\times$10$^{17}$ Mx.  During the cancellation the positive and negative polarities lose 0.7$\times$10$^{17}$ Mx and 1.9$\times$10$^{17}$ Mx, respectively. The average cancellation rate was 3.7$\times$10$^{14}$ Mx s$^{-1}$. Within about 2 minutes of cancelling there was a 0.6 km~s$^{-1}$ increase in submergence speed, then a gradual oscillation and decrease to around 0 km~s$^{-1}$. 

\subsection{Analysis of ROI\_08}

\begin{figure*}[ht!]
\includegraphics[width=1.0\linewidth]{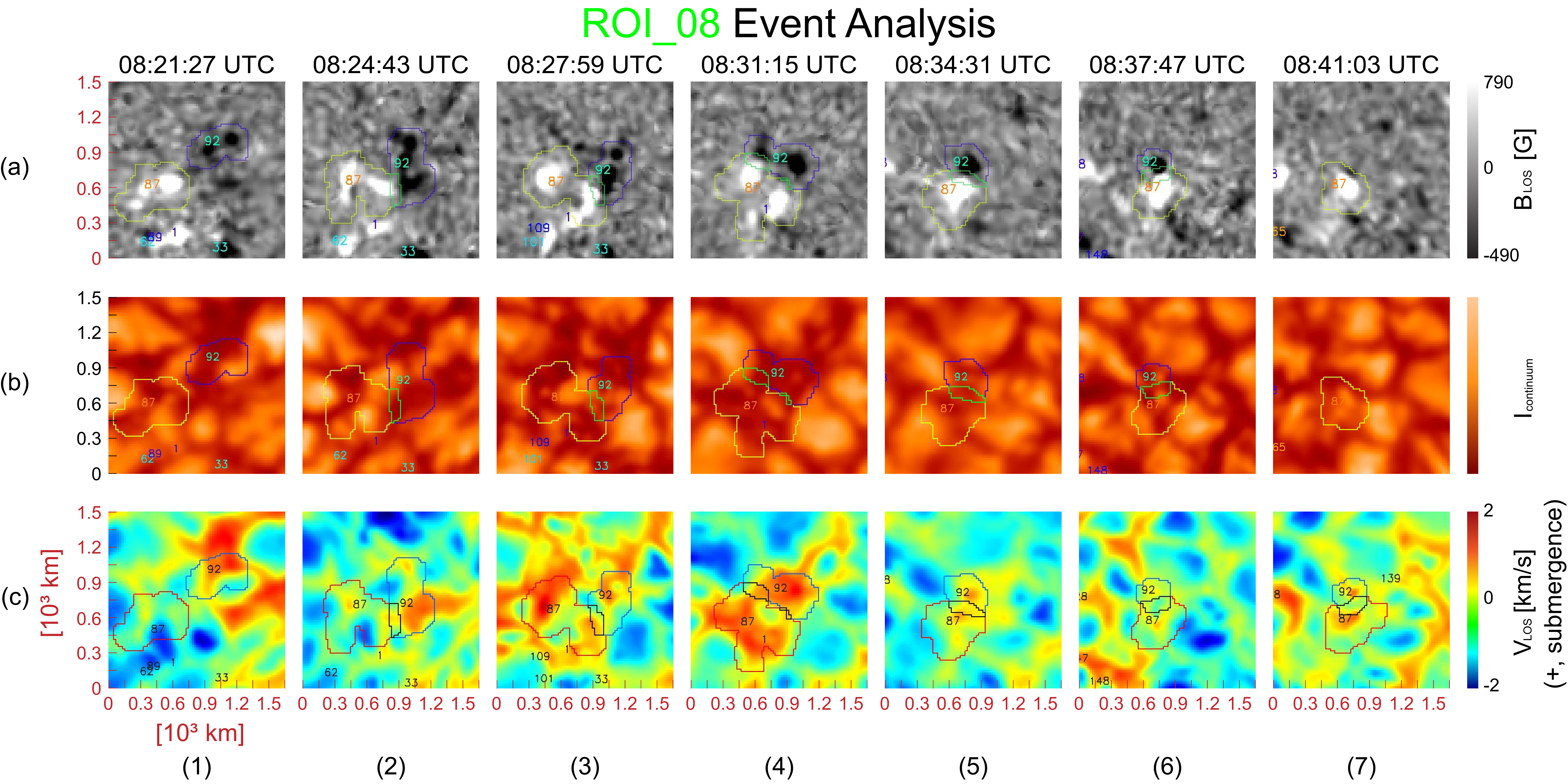}
\caption{Evolution of B$_\mathrm{LOS}$ (top series), 630.1 nm intensity (middle series) and $V_\mathrm{LOS}$ (bottom series). ROI\_{08} in the time series begins with panel 1 (time immediately before PIL is defined) and progresses to panel 7 (time immediately after PIL is no longer defined).
\label{fig:roi08fulldataset}}
\end{figure*}

\begin{figure}[ht!]
\centering
\includegraphics[width=.5\linewidth]{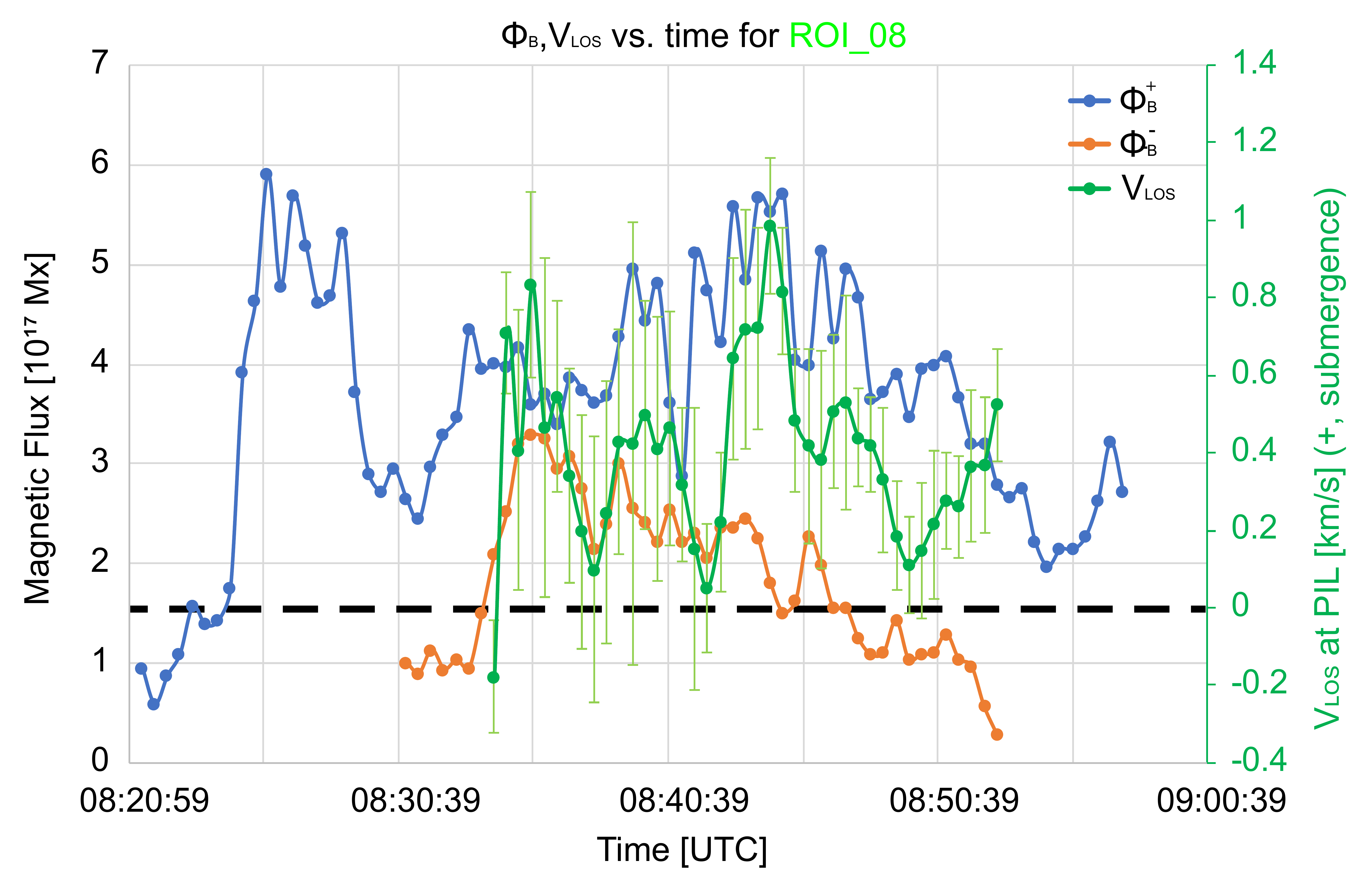}
\caption{Evolution of LOS magnetic flux and PIL mean Doppler velocity, $\Phi_\mathrm{B}$ and $V_\mathrm{LOS}$, in ROI\_08. Refer to the caption of Figure~\ref{fig:roi03plot} for an explanation of the legend and other graphical details
\label{fig:roi08plot}}
\end{figure}

In ROI$\_08$ we see cancellation occurring by examining the top row of Figure~\ref{fig:roi08fulldataset}. Examining continuum imagery, we see the cancellation occurred in an intergranular lane. Positive polarity labeled 87 and negative polarity labeled 92, seen in Figure~\ref{fig:roi08fulldataset} begin cancelling at 08:21:27 UTC and the event lasts until 08:41:03 UTC. 
$\mathrm \Phi_{B,i}^{-}$ was 2.1$\times$10$^{17}$ Mx and $\mathrm \Phi_{B,i}^{+}$ was 4.02$\times$10$^{17}$ Mx 
and during the cancellation the positive and 
negative polarities lose 1.2$\times$10$^{17}$ Mx and 1.8$\times$10$^{17}$ Mx, respectively. 
About 50\% of the initial bipole magnetic flux was lost during the cancellation event. 
The average cancellation rate through the event was 2.5$\times$10$^{14}$ Mx s$^{-1}$. 
Visual inspection of the magnetograms in Figure~\ref{fig:roi08fulldataset}, 
reveals that both the positive and negative polarities lose apparent size as they cancel, with the negative polarity falling below the instrument noise level in panel a7. Figure~\ref{fig:roi08plot} shows that within roughly 2 minutes after first cancelling the Doppler velocity at the PIL jumps from -0.2 km~s$^{-1}$ to 0.8 km~s$^{-1}$. Following that initial increase the Doppler velocity is relatively variable but still higher than the initial -0.2 km~s$^{-1}$ detected.

\subsection{Analysis of ROI\_11}

\begin{figure*}[ht!]
\includegraphics[width=1.0\linewidth]{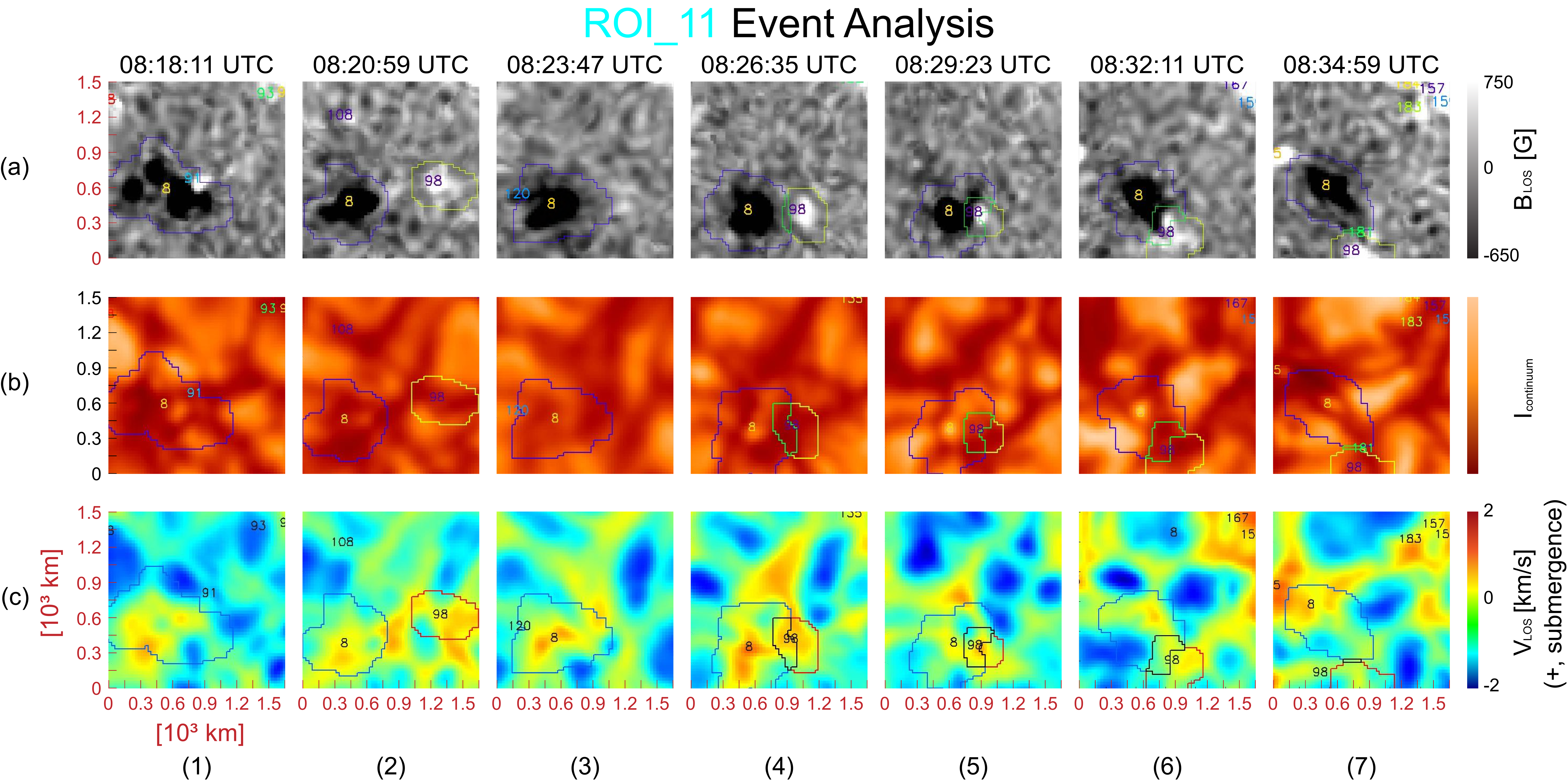}
\caption{Evolution of B$_\mathrm{LOS}$ (top series), 630.1 nm intensity (middle series) and $V_\mathrm{LOS}$ (bottom series). $ROI_{11}$ in the time series begins with panel 1 (time immediately before PIL is defined) and progresses to panel 7 (time immediately after PIL is no longer defined).
\label{fig:roi11fulldataset}}
\end{figure*}

\begin{figure}[ht!]
\centering
\includegraphics[width=.5\linewidth]{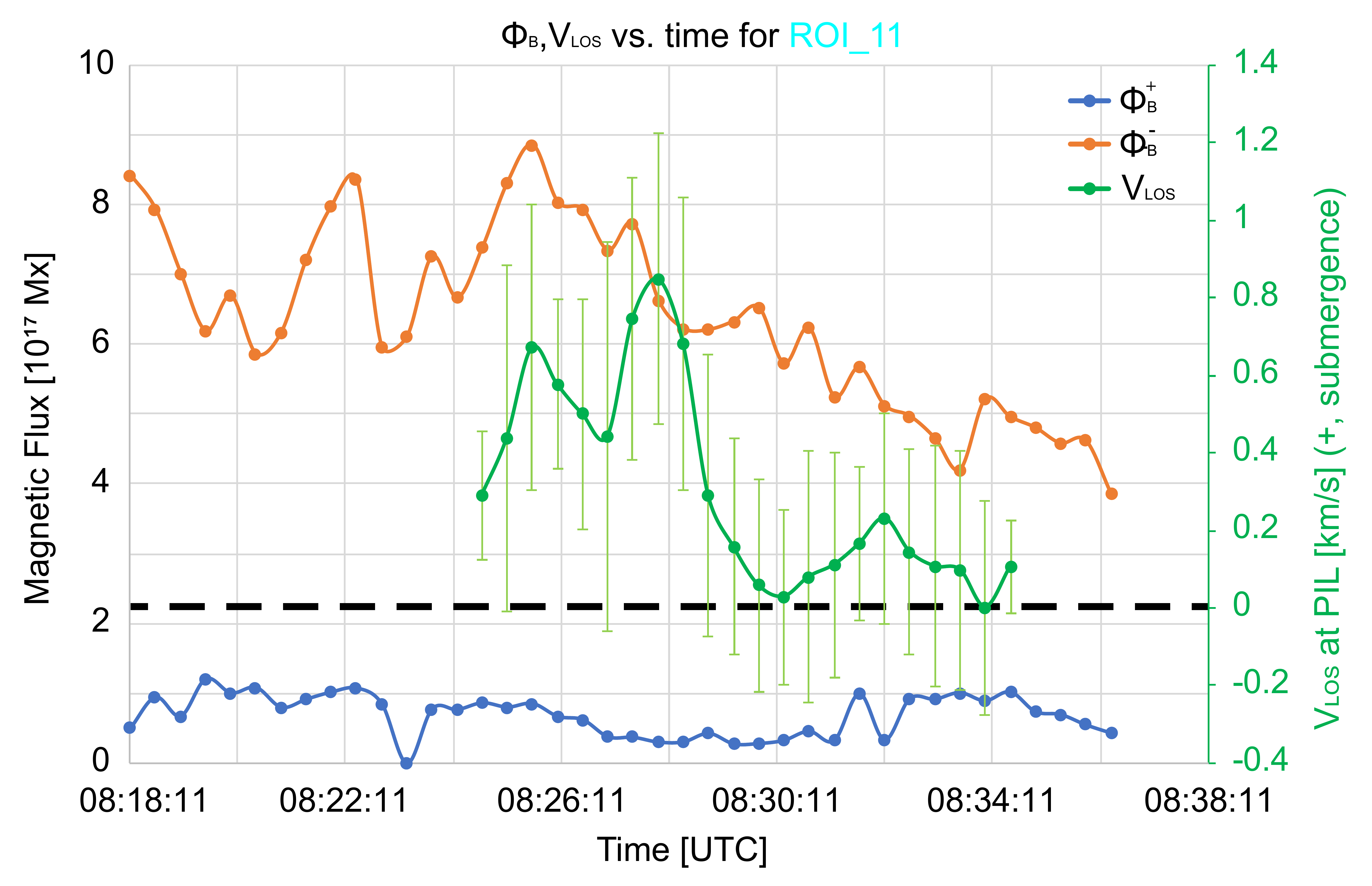}
\caption{Evolution of LOS magnetic flux and PIL mean Doppler velocity, $\Phi_\mathrm{B}$ and $V_\mathrm{LOS}$, in ROI\_11. Refer to the caption of Figure~\ref{fig:roi03plot} for an explanation of the legend and other graphical details.
\label{fig:roi11plot}}
\end{figure}

ROI\_11 involved a small positive polarity ($0.9\times10^{17}$ Mx) and a large negative polarity ($7.4\times10^{17}$ Mx) that cancelled over the course of around 18 minutes. The cancellation began in a downflow of 0.3 km~s$^{-1}$ and the submergence speed peaked at around 0.8 km~s$^{-1}$, the same time as the cancellation rate peaked ($R_\mathrm{peak}=27\times10^{14}$ Mx). The positive polarity gained a small amount of flux during the event, around $0.1\times10^{17}$ Mx while the negative polarity's flux decreased almost 30\% or $2.4\times10^{17}$ Mx. These values can be inferred from inspection of Figure~\ref{fig:roi11plot}. Analyzing the $B_\mathrm{LOS}$, $I_\mathrm{continuum}$, and $V_\mathrm{LOS}$ images in Figure~\ref{fig:roi11fulldataset}, we can see in continuum imagery that the cancellation begins in an intergranular lane (dark patch). This is supported by the starting $V_\mathrm{LOS}$ of 0.3 km~s$^{-1}$. In $B_\mathrm{LOS}$ we can see the positive polarity patch labeled 8 and negative polarity patch labeled 98 interacting throughout the time series and eventually the positive polarity moves out of the ROI in the last time step.
In panel 4c we see the greatest intensification of the $V_\mathrm{LOS}$. In panel 3a we see the positive polarity briefly dip below the noise threshold then it is re-labeled in panel 4a.

\subsection{Analysis of ROI\_23}

\begin{figure*}[ht!]
\includegraphics[width=1.0\linewidth]{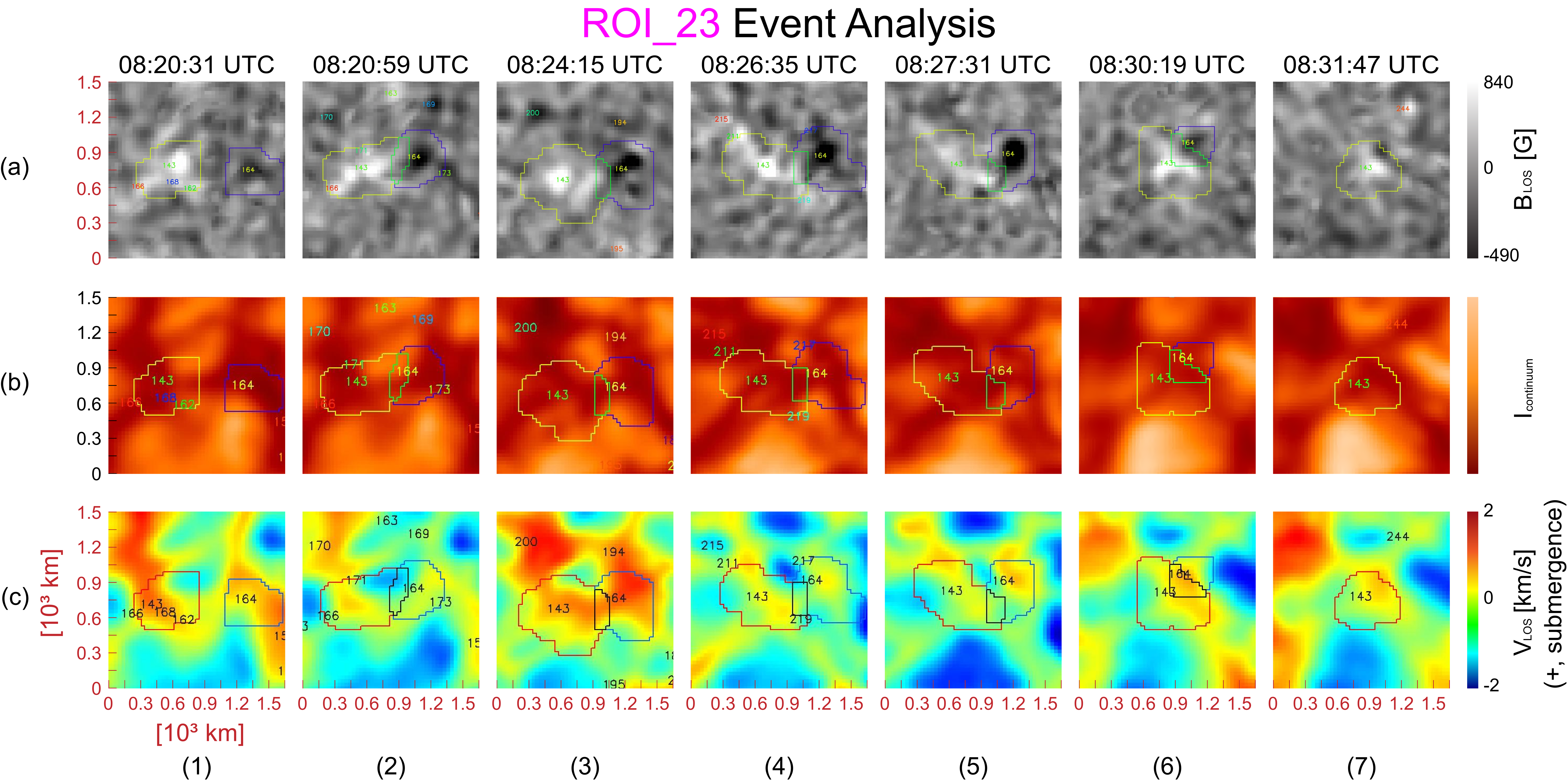}
\caption{Evolution of B$_\mathrm{LOS}$ (top series), 630.1 nm line core (middle series) and $V_\mathrm{LOS}$ (bottom series). $ROI_{23}$ in the time series begins with panel 1 (time immediately before PIL is defined) and progresses to panel 7 (time immediately after PIL is no longer defined).
\label{fig:roi23fulldataset}}
\end{figure*}

\begin{figure}[ht!]
\centering
\includegraphics[width=.5\linewidth]{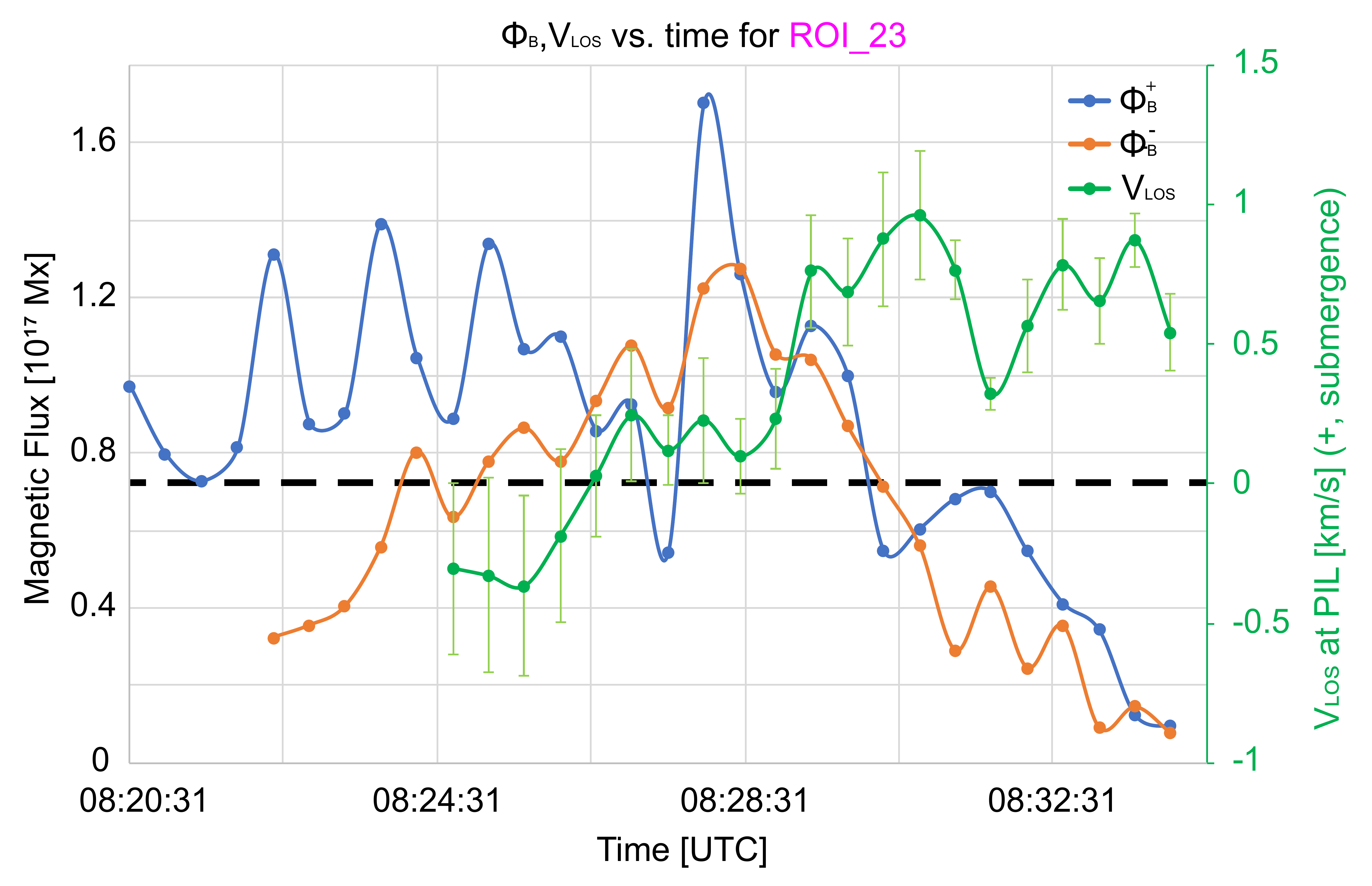}
\caption{Evolution of LOS magnetic flux and PIL mean Doppler velocity, $\Phi_\mathrm{B}$ and $V_\mathrm{LOS}$, in ROI\_23. Refer to the caption of Figure~\ref{fig:roi03plot} for an explanation of the legend and other graphical details.
\label{fig:roi23plot}}
\end{figure}

ROI\_23 involves a large positive polarity ($1.0\times10^{17}$ Mx) and a small negative polarity ($0.5\times10^{17}$ Mx). The cancellation lasted 28 minutes and reached a maximum Doppler velocity of $V_\mathrm{LOS}$ = 1.1 km~s$^{-1}$. This event is different from most in that it begins in an upflow of roughly 0.3 km~s$^{-1}$ and ends in a downflow of roughly 1.1 km~$^{-1}$ meaning that the $\Delta V_\mathrm{LOS}$ of this event is large, $\approx$ 1.4 km~s$^{-1}$.  Both Figure~\ref{fig:roi23fulldataset} and the plot in Figure~\ref{fig:roi23plot} show that this is a very clear case of cancellation ending in a downflow region in an intergranular lane. 

\clearpage
\bibliography{apj-jour,master}

\begin{thebibliography}{}
\expandafter\ifx\csname natexlab\endcsname\relax\def\natexlab#1{#1}\fi
\providecommand{\url}[1]{\href{#1}{#1}}
\providecommand{\dodoi}[1]{doi:~\href{http://doi.org/#1}{\nolinkurl{#1}}}
\providecommand{\doeprint}[1]{\href{http://ascl.net/#1}{\nolinkurl{http://ascl.net/#1}}}
\providecommand{\doarXiv}[1]{\href{https://arxiv.org/abs/#1}{\nolinkurl{https://arxiv.org/abs/#1}}}

\bibitem[{{Bellot Rubio} \& {Orozco Su{\'a}rez}(2019)}]{bellotrubio2019}
{Bellot Rubio}, L., \& {Orozco Su{\'a}rez}, D. 2019, Living Reviews in Solar
  Physics, 16, 1, \dodoi{10.1007/s41116-018-0017-1}

\bibitem[{{Bellot Rubio} \& {Orozco Su{\'a}rez}(2012)}]{introduction_6}
{Bellot Rubio}, L.~R., \& {Orozco Su{\'a}rez}, D. 2012, \apj, 757, 19,
  \dodoi{10.1088/0004-637X/757/1/19}

\bibitem[{{Chae} {et~al.}(2004){Chae}, {Moon}, \&
  {Pevtsov}}]{downflows_omegaloop}
{Chae}, J., {Moon}, Y.-J., \& {Pevtsov}, A.~A. 2004, \apjl, 602, L65,
  \dodoi{10.1086/382222}

\bibitem[{{Chae} {et~al.}(2002){Chae}, {Moon}, {Wang}, \&
  {Yun}}]{CancellationRates}
{Chae}, J., {Moon}, Y.-J., {Wang}, H., \& {Yun}, H.~S. 2002, \solphys, 207, 73,
  \dodoi{10.1023/A:1015534219066}

\bibitem[{{Chintzoglou} {et~al.}(2019){Chintzoglou}, {Zhang}, {Cheung}, \&
  {Kazachenko}}]{Chintzoglou2019}
{Chintzoglou}, G., {Zhang}, J., {Cheung}, M. C.~M., \& {Kazachenko}, M. 2019,
  \apj, 871, 67, \dodoi{10.3847/1538-4357/aaef30}

\bibitem[{{Criscuoli} \& {Foukal}(2017)}]{criscuoli2017}
{Criscuoli}, S., \& {Foukal}, P. 2017, \apj, 835, 99,
  \dodoi{10.3847/1538-4357/835/1/99}

\bibitem[{{Cristaldi} \& {Ermolli}(2017)}]{irradiance}
{Cristaldi}, A., \& {Ermolli}, I. 2017, \apj, 841, 115,
  \dodoi{10.3847/1538-4357/aa713c}

\bibitem[{{Danilovic} {et~al.}(2010){Danilovic}, {Beeck}, {Pietarila},
  {Sch{\"u}ssler}, {Solanki}, {Mart{\'\i}nez Pillet}, {Bonet}, {del Toro
  Iniesta}, {Domingo}, {Barthol}, {Berkefeld}, {Gandorfer}, {Kn{\"o}lker},
  {Schmidt}, \& {Title}}]{qsfieldresolved}
{Danilovic}, S., {Beeck}, B., {Pietarila}, A., {et~al.} 2010, \apjl, 723, L149,
  \dodoi{10.1088/2041-8205/723/2/L149}

\bibitem[{{de la Cruz Rodr{\'\i}guez} {et~al.}(2015){de la Cruz
  Rodr{\'\i}guez}, {L{\"o}fdahl}, {S{\"u}tterlin}, {Hillberg}, \& {Rouppe van
  der Voort}}]{CRISPRED}
{de la Cruz Rodr{\'\i}guez}, J., {L{\"o}fdahl}, M.~G., {S{\"u}tterlin}, P.,
  {Hillberg}, T., \& {Rouppe van der Voort}, L. 2015, \aap, 573, A40,
  \dodoi{10.1051/0004-6361/201424319}

\bibitem[{{De Pontieu} {et~al.}(2014){De Pontieu}, {Title}, {Lemen}, {Kushner},
  {Akin}, {Allard}, {Berger}, {Boerner}, {Cheung}, {Chou}, {Drake}, {Duncan},
  {Freeland}, {Heyman}, {Hoffman}, {Hurlburt}, {Lindgren}, {Mathur}, {Rehse},
  {Sabolish}, {Seguin}, {Schrijver}, {Tarbell}, {W{\"u}lser}, {Wolfson},
  {Yanari}, {Mudge}, {Nguyen-Phuc}, {Timmons}, {van Bezooijen}, {Weingrod},
  {Brookner}, {Butcher}, {Dougherty}, {Eder}, {Knagenhjelm}, {Larsen},
  {Mansir}, {Phan}, {Boyle}, {Cheimets}, {DeLuca}, {Golub}, {Gates}, {Hertz},
  {McKillop}, {Park}, {Perry}, {Podgorski}, {Reeves}, {Saar}, {Testa}, {Tian},
  {Weber}, {Dunn}, {Eccles}, {Jaeggli}, {Kankelborg}, {Mashburn}, {Pust},
  {Springer}, {Carvalho}, {Kleint}, {Marmie}, {Mazmanian}, {Pereira}, {Sawyer},
  {Strong}, {Worden}, {Carlsson}, {Hansteen}, {Leenaarts}, {Wiesmann},
  {Aloise}, {Chu}, {Bush}, {Scherrer}, {Brekke}, {Martinez-Sykora}, {Lites},
  {McIntosh}, {Uitenbroek}, {Okamoto}, {Gummin}, {Auker}, {Jerram}, {Pool}, \&
  {Waltham}}]{IRIS}
{De Pontieu}, B., {Title}, A.~M., {Lemen}, J.~R., {et~al.} 2014, \solphys, 289,
  2733, \dodoi{10.1007/s11207-014-0485-y}

\bibitem[{{de Wijn} {et~al.}(2008){de Wijn}, {Lites}, {Berger}, {Frank},
  {Tarbell}, \& {Ishikawa}}]{Internetwork_3_Support_5}
{de Wijn}, A.~G., {Lites}, B.~W., {Berger}, T.~E., {et~al.} 2008, \apj, 684,
  1469, \dodoi{10.1086/590237}

\bibitem[{{DeForest} {et~al.}(2007){DeForest}, {Hagenaar}, {Lamb}, {Parnell},
  \& {Welsch}}]{threshold}
{DeForest}, C.~E., {Hagenaar}, H.~J., {Lamb}, D.~A., {Parnell}, C.~E., \&
  {Welsch}, B.~T. 2007, \apj, 666, 576, \dodoi{10.1086/518994}

\bibitem[{{del Toro Iniesta}(2007)}]{deltoro2007}
{del Toro Iniesta}, J.~C. 2007, {Introduction to Spectropolarimetry}

\bibitem[{{del Toro Iniesta} \& {Ruiz Cobo}(2016)}]{editedbib1}
{del Toro Iniesta}, J.~C., \& {Ruiz Cobo}, B. 2016, Living Reviews in Solar
  Physics, 13, 4, \dodoi{10.1007/s41116-016-0005-2}

\bibitem[{{Denker} \& {Tritschler}(2009)}]{prominences}
{Denker}, C., \& {Tritschler}, A. 2009, in IAU Symposium, Vol. 259, Cosmic
  Magnetic Fields: From Planets, to Stars and Galaxies, ed. K.~G.
  {Strassmeier}, A.~G. {Kosovichev}, \& J.~E. {Beckman}, 223--224,
  \dodoi{10.1017/S1743921309030476}

\bibitem[{{Dravins} {et~al.}(1981){Dravins}, {Lindegren}, \&
  {Nordlund}}]{salvocitation2}
{Dravins}, D., {Lindegren}, L., \& {Nordlund}, A. 1981, \aap, 96, 345

\bibitem[{{Faurobert} {et~al.}(2016){Faurobert}, {Balasubramanian}, \&
  {Ricort}}]{faurobert2016}
{Faurobert}, M., {Balasubramanian}, R., \& {Ricort}, G. 2016, \aap, 595, A71,
  \dodoi{10.1051/0004-6361/201527797}

\bibitem[{{Gosic} {et~al.}(2017){Gosic}, {de la Cruz Rodriguez}, {De Pontieu},
  {Bellot Rubio}, {Esteban Pozuelo}, \&
  {Ortiz-Carbonell}}]{Chromosphere_1_Support_1}
{Gosic}, M., {de la Cruz Rodriguez}, J., {De Pontieu}, B., {et~al.} 2017, in
  AGU Fall Meeting Abstracts, Vol. 2017, SH41C--02

\bibitem[{{Gosic} {et~al.}(2018){Gosic}, {De Pontieu}, \& {Bellot
  Rubio}}]{Chromosphere_1_Support_2}
{Gosic}, M., {De Pontieu}, B., \& {Bellot Rubio}, L.~R. 2018, in 42nd COSPAR
  Scientific Assembly, Vol.~42, E2.3--2--18

\bibitem[{{Gosic} {et~al.}(2012){Gosic}, {Katsukawa}, {Bellot Rubio}, \&
  {Orozco Suarez}}]{InitialFluxMeasurementGosic}
{Gosic}, M., {Katsukawa}, Y., {Bellot Rubio}, L., \& {Orozco Suarez}, D. 2012,
  in 39th COSPAR Scientific Assembly, Vol.~39, 657

\bibitem[{{Go{\v{s}}i{\'c}} {et~al.}(2016){Go{\v{s}}i{\'c}}, {Bellot Rubio},
  {del Toro Iniesta}, {Orozco Su{\'a}rez}, \& {Katsukawa}}]{Internetwork_1}
{Go{\v{s}}i{\'c}}, M., {Bellot Rubio}, L.~R., {del Toro Iniesta}, J.~C.,
  {Orozco Su{\'a}rez}, D., \& {Katsukawa}, Y. 2016, \apj, 820, 35,
  \dodoi{10.3847/0004-637X/820/1/35}

\bibitem[{{Go{\v{s}}i{\'c}} {et~al.}(2014){Go{\v{s}}i{\'c}}, {Bellot Rubio},
  {Orozco Su{\'a}rez}, {Katsukawa}, \& {del Toro Iniesta}}]{introduction_7}
{Go{\v{s}}i{\'c}}, M., {Bellot Rubio}, L.~R., {Orozco Su{\'a}rez}, D.,
  {Katsukawa}, Y., \& {del Toro Iniesta}, J.~C. 2014, \apj, 797, 49,
  \dodoi{10.1088/0004-637X/797/1/49}

\bibitem[{{Go{\v{s}}i{\'c}} {et~al.}(2018){Go{\v{s}}i{\'c}}, {de la Cruz
  Rodr{\'\i}guez}, {De Pontieu}, {Bellot Rubio}, {Carlsson}, {Esteban Pozuelo},
  {Ortiz}, \& {Polito}}]{Chromosphere_1}
{Go{\v{s}}i{\'c}}, M., {de la Cruz Rodr{\'\i}guez}, J., {De Pontieu}, B.,
  {et~al.} 2018, \apj, 857, 48, \dodoi{10.3847/1538-4357/aab1f0}

\bibitem[{{Gošić}(2015)}]{CancellationLifeTimeGosic}
{Gošić}, M. 2015, {The solar internetwork. PhD thesis, Universidad de
  Granada, Spain}

\bibitem[{{Guglielmino} {et~al.}(2012){Guglielmino}, {Mart{\'\i}nez Pillet},
  {Bonet}, {del Toro Iniesta}, {Bellot Rubio}, {Solanki}, {Schmidt},
  {Gandorfer}, {Barthol}, \& {Kn{\"o}lker}}]{Cancellation_3}
{Guglielmino}, S.~L., {Mart{\'\i}nez Pillet}, V., {Bonet}, J.~A., {et~al.}
  2012, \apj, 745, 160, \dodoi{10.1088/0004-637X/745/2/160}

\bibitem[{{Harvey} {et~al.}(1999){Harvey}, {Jones}, {Schrijver}, \&
  {Penn}}]{fluxsubmergence}
{Harvey}, K.~L., {Jones}, H.~P., {Schrijver}, C.~J., \& {Penn}, M.~J. 1999,
  \solphys, 190, 35, \dodoi{10.1023/A:1005237719407}

\bibitem[{{Iida} {et~al.}(2012){Iida}, {Hagenaar}, \&
  {Yokoyama}}]{supergranularmotion}
{Iida}, Y., {Hagenaar}, H.~J., \& {Yokoyama}, T. 2012, \apj, 752, 149,
  \dodoi{10.1088/0004-637X/752/2/149}

\bibitem[{{Kaithakkal} \& {Solanki}(2019{\natexlab{a}})}]{Cancellation_1}
{Kaithakkal}, A.~J., \& {Solanki}, S.~K. 2019{\natexlab{a}}, \aap, 622, A200,
  \dodoi{10.1051/0004-6361/201833770}

\bibitem[{{Kaithakkal} \& {Solanki}(2019{\natexlab{b}})}]{Internetwork_3}
---. 2019{\natexlab{b}}, \aap, 622, A200, \dodoi{10.1051/0004-6361/201833770}

\bibitem[{{Keys} {et~al.}(2011){Keys}, {Mathioudakis}, {Jess}, {Shelyag},
  {Crockett}, {Christian}, \& {Keenan}}]{keys2011}
{Keys}, P.~H., {Mathioudakis}, M., {Jess}, D.~B., {et~al.} 2011, \apjl, 740,
  L40, \dodoi{10.1088/2041-8205/740/2/L40}

\bibitem[{{Kontogiannis} {et~al.}(2020){Kontogiannis}, {Tsiropoula},
  {Tziotziou}, {Gontikakis}, {Kuckein}, {Verma}, \& {Denker}}]{Chromosphere_2}
{Kontogiannis}, I., {Tsiropoula}, G., {Tziotziou}, K., {et~al.} 2020, \aap,
  633, A67, \dodoi{10.1051/0004-6361/201936778}

\bibitem[{{Kubo} {et~al.}(2010){Kubo}, {Low}, \& {Lites}}]{Kubo}
{Kubo}, M., {Low}, B.~C., \& {Lites}, B.~W. 2010, \apj, 712, 1321,
  \dodoi{10.1088/0004-637X/712/2/1321}

\bibitem[{{Lamb} {et~al.}(2008){Lamb}, {DeForest}, {Hagenaar}, {Parnell}, \&
  {Welsch}}]{lambetal}
{Lamb}, D.~A., {DeForest}, C.~E., {Hagenaar}, H.~J., {Parnell}, C.~E., \&
  {Welsch}, B.~T. 2008, \apj, 674, 520, \dodoi{10.1086/524372}

\bibitem[{{Lamb} {et~al.}(2013){Lamb}, {Howard}, {DeForest}, {Parnell}, \&
  {Welsch}}]{evidence1}
{Lamb}, D.~A., {Howard}, T.~A., {DeForest}, C.~E., {Parnell}, C.~E., \&
  {Welsch}, B.~T. 2013, \apj, 774, 127, \dodoi{10.1088/0004-637X/774/2/127}

\bibitem[{{Litvinenko} {et~al.}(2007){Litvinenko}, {Chae}, \&
  {Park}}]{SpecificCancellationRate}
{Litvinenko}, Y.~E., {Chae}, J., \& {Park}, S.-Y. 2007, \apj, 662, 1302,
  \dodoi{10.1086/518115}

\bibitem[{{Livi} {et~al.}(1985){Livi}, {Wang}, \& {Martin}}]{introduction_8}
{Livi}, S.~H.~B., {Wang}, J., \& {Martin}, S.~F. 1985, Australian Journal of
  Physics, 38, 855, \dodoi{10.1071/PH850855}

\bibitem[{{Livingston} \& {Harvey}(1971)}]{weaksignals}
{Livingston}, W., \& {Harvey}, J. 1971, in Solar Magnetic Fields, ed.
  R.~{Howard}, Vol.~43, 51

\bibitem[{{Livingston} \& {Harvey}(1975)}]{introduction_4}
{Livingston}, W.~C., \& {Harvey}, J. 1975, in \baas, Vol.~7, 346

\bibitem[{{Martin} {et~al.}(1985){Martin}, {Livi}, \& {Wang}}]{introduction_9}
{Martin}, S.~F., {Livi}, S.~H.~B., \& {Wang}, J. 1985, Australian Journal of
  Physics, 38, 929, \dodoi{10.1071/PH850929}

\bibitem[{{Mart{\'\i}nez Gonz{\'a}lez} \& {Bellot
  Rubio}(2009)}]{introduction_3}
{Mart{\'\i}nez Gonz{\'a}lez}, M.~J., \& {Bellot Rubio}, L.~R. 2009, \apj, 700,
  1391, \dodoi{10.1088/0004-637X/700/2/1391}

\bibitem[{{Nisenson} {et~al.}(2003){Nisenson}, {van Ballegooijen}, {de Wijn},
  \& {S{\"u}tterlin}}]{rmsvelocity}
{Nisenson}, P., {van Ballegooijen}, A.~A., {de Wijn}, A.~G., \&
  {S{\"u}tterlin}, P. 2003, \apj, 587, 458, \dodoi{10.1086/368067}

\bibitem[{{Oba} {et~al.}(2017){Oba}, {Iida}, \& {Shimizu}}]{intergranularflow}
{Oba}, T., {Iida}, Y., \& {Shimizu}, T. 2017, \apj, 836, 40,
  \dodoi{10.3847/1538-4357/836/1/40}

\bibitem[{{Panesar} {et~al.}(2016){Panesar}, {Sterling}, {Moore}, \&
  {Chakrapani}}]{coronaljets}
{Panesar}, N.~K., {Sterling}, A.~C., {Moore}, R.~L., \& {Chakrapani}, P. 2016,
  \apjl, 832, L7, \dodoi{10.3847/2041-8205/832/1/L7}

\bibitem[{{Park} {et~al.}(2009){Park}, {Chae}, \&
  {Litvinenko}}]{Cancellation_4}
{Park}, S., {Chae}, J., \& {Litvinenko}, Y.~E. 2009, \apjl, 704, L71,
  \dodoi{10.1088/0004-637X/704/1/L71}

\bibitem[{{Rast} {et~al.}(2021){Rast}, {Bello Gonz{\'a}lez}, {Bellot Rubio},
  {Cao}, {Cauzzi}, {Deluca}, {de Pontieu}, {Fletcher}, {Gibson}, {Judge},
  {Katsukawa}, {Kazachenko}, {Khomenko}, {Landi}, {Mart{\'\i}nez Pillet},
  {Petrie}, {Qiu}, {Rachmeler}, {Rempel}, {Schmidt}, {Scullion}, {Sun},
  {Welsch}, {Andretta}, {Antolin}, {Ayres}, {Balasubramaniam}, {Ballai},
  {Berger}, {Bradshaw}, {Campbell}, {Carlsson}, {Casini}, {Centeno}, {Cranmer},
  {Criscuoli}, {Deforest}, {Deng}, {Erd{\'e}lyi}, {Fedun}, {Fischer},
  {Gonz{\'a}lez Manrique}, {Hahn}, {Harra}, {Henriques}, {Hurlburt}, {Jaeggli},
  {Jafarzadeh}, {Jain}, {Jefferies}, {Keys}, {Kowalski}, {Kuckein}, {Kuhn},
  {Kuridze}, {Liu}, {Liu}, {Longcope}, {Mathioudakis}, {McAteer}, {McIntosh},
  {McKenzie}, {Miralles}, {Morton}, {Muglach}, {Nelson}, {Panesar}, {Parenti},
  {Parnell}, {Poduval}, {Reardon}, {Reep}, {Schad}, {Schmit}, {Sharma},
  {Socas-Navarro}, {Srivastava}, {Sterling}, {Suematsu}, {Tarr}, {Tiwari},
  {Tritschler}, {Verth}, {Vourlidas}, {Wang}, {Wang}, {NSO and DKIST Project},
  {DKIST Instrument Scientists}, {DKIST Science Working Group}, \& {DKIST
  Critical Science Plan Community}}]{Rast2021}
{Rast}, M.~P., {Bello Gonz{\'a}lez}, N., {Bellot Rubio}, L., {et~al.} 2021,
  \solphys, 296, 70, \dodoi{10.1007/s11207-021-01789-2}

\bibitem[{{Rees} \& {Semel}(1979)}]{rees1979}
{Rees}, D.~E., \& {Semel}, M.~D. 1979, \aap, 74, 1

\bibitem[{{Rempel}(2014)}]{rempel2014}
{Rempel}, M. 2014, \apj, 789, 132, \dodoi{10.1088/0004-637X/789/2/132}

\bibitem[{{Rempel}(2020)}]{rempel2020}
---. 2020, \apj, 894, 140, \dodoi{10.3847/1538-4357/ab8633}

\bibitem[{{S{\'a}nchez Almeida}(2004)}]{introduction_1}
{S{\'a}nchez Almeida}, J. 2004, Astronomical Society of the Pacific Conference
  Series, Vol. 325, {The Magnetism of the Very Quiet Sun}, ed. T.~{Sakurai} \&
  T.~{Sekii}, 115

\bibitem[{{Scharmer} {et~al.}(2003{\natexlab{a}}){Scharmer}, {Bjelksjo},
  {Korhonen}, {Lindberg}, \& {Petterson}}]{scharmer2003}
{Scharmer}, G.~B., {Bjelksjo}, K., {Korhonen}, T.~K., {Lindberg}, B., \&
  {Petterson}, B. 2003{\natexlab{a}}, in Society of Photo-Optical
  Instrumentation Engineers (SPIE) Conference Series, Vol. 4853, Innovative
  Telescopes and Instrumentation for Solar Astrophysics, ed. S.~L. {Keil} \&
  S.~V. {Avakyan}, 341--350, \dodoi{10.1117/12.460377}

\bibitem[{{Scharmer} {et~al.}(2003{\natexlab{b}}){Scharmer}, {Dettori},
  {L{\"o}fdahl}, \& {Shand}}]{adaptiveoptics}
{Scharmer}, G.~B., {Dettori}, P.~M., {L{\"o}fdahl}, M.~G., \& {Shand}, M.
  2003{\natexlab{b}}, in \procspie, Vol. 4853, Innovative Telescopes and
  Instrumentation for Solar Astrophysics, ed. S.~L. {Keil} \& S.~V. {Avakyan},
  370--380, \dodoi{10.1117/12.460387}

\bibitem[{{Scharmer} {et~al.}(2008){Scharmer}, {Narayan}, {Hillberg}, {de la
  Cruz Rodriguez}, {L{\"o}fdahl}, {Kiselman}, {S{\"u}tterlin}, {van Noort}, \&
  {Lagg}}]{CRISP}
{Scharmer}, G.~B., {Narayan}, G., {Hillberg}, T., {et~al.} 2008, \apjl, 689,
  L69, \dodoi{10.1086/595744}

\bibitem[{{Schmieder} {et~al.}(2002){Schmieder}, {Pariat}, {Aulanier},
  {Georgoulis}, {Rust}, \& {Bernasconi}}]{ellermanbombs}
{Schmieder}, B., {Pariat}, E., {Aulanier}, G., {et~al.} 2002, in ESA Special
  Publication, Vol.~2, Solar Variability: From Core to Outer Frontiers, ed.
  A.~{Wilson}, 911--914

\bibitem[{{Schrijver} {et~al.}(1997){Schrijver}, {Title}, {van Ballegooijen},
  {Hagenaar}, \& {Shine}}]{introduction_10}
{Schrijver}, C.~J., {Title}, A.~M., {van Ballegooijen}, A.~A., {Hagenaar},
  H.~J., \& {Shine}, R.~A. 1997, \apj, 487, 424, \dodoi{10.1086/304581}

\bibitem[{{Stangalini} {et~al.}(2015){Stangalini}, {Giannattasio}, \&
  {Jafarzadeh}}]{kinkwaves1}
{Stangalini}, M., {Giannattasio}, F., \& {Jafarzadeh}, S. 2015, \aap, 577, A17,
  \dodoi{10.1051/0004-6361/201425273}

\bibitem[{{Stangalini} {et~al.}(2017){Stangalini}, {Giannattasio},
  {Erd{\'e}lyi}, {Jafarzadeh}, {Consolini}, {Criscuoli}, {Ermolli},
  {Guglielmino}, \& {Zuccarello}}]{kinkwaves2}
{Stangalini}, M., {Giannattasio}, F., {Erd{\'e}lyi}, R., {et~al.} 2017, \apj,
  840, 19, \dodoi{10.3847/1538-4357/aa6c5e}

\bibitem[{{Thornton} \& {Parnell}(2011)}]{introduction_2}
{Thornton}, L.~M., \& {Parnell}, C.~E. 2011, \solphys, 269, 13,
  \dodoi{10.1007/s11207-010-9656-7}

\bibitem[{{Uitenbroek}(2003)}]{uitenbroek2003}
{Uitenbroek}, H. 2003, \apj, 592, 1225, \dodoi{10.1086/375736}

\bibitem[{{van Noort} {et~al.}(2005){van Noort}, {Rouppe van der Voort}, \&
  {L{\"o}fdahl}}]{MOMFBD}
{van Noort}, M., {Rouppe van der Voort}, L., \& {L{\"o}fdahl}, M.~G. 2005,
  \solphys, 228, 191, \dodoi{10.1007/s11207-005-5782-z}

\bibitem[{{Viavattene} {et~al.}(2021){Viavattene}, {Murabito}, {Guglielmino},
  {Ermolli}, {Consolini}, {Giorgi}, \& {Jafarzadeh}}]{salvocitation}
{Viavattene}, G., {Murabito}, M., {Guglielmino}, S.~L., {et~al.} 2021, Entropy,
  23, 413, \dodoi{10.3390/e23040413}

\bibitem[{{Wang} {et~al.}(1996){Wang}, {Shi}, \& {Martin}}]{introduction_12}
{Wang}, J., {Shi}, Z., \& {Martin}, S.~F. 1996, \aap, 316, 201

\bibitem[{{Welsch} {et~al.}(2004){Welsch}, {Fisher}, {Abbett}, \&
  {Regnier}}]{YAFTA_1}
{Welsch}, B.~T., {Fisher}, G.~H., {Abbett}, W.~P., \& {Regnier}, S. 2004, \apj,
  610, 1148, \dodoi{10.1086/421767}

\bibitem[{{Yang} {et~al.}(2009){Yang}, {Zhang}, \& {Borrero}}]{granularmotions}
{Yang}, S., {Zhang}, J., \& {Borrero}, J.~M. 2009, \apj, 703, 1012,
  \dodoi{10.1088/0004-637X/703/1/1012}

\bibitem[{{Yardley} {et~al.}(2016){Yardley}, {Green}, {Williams}, {van
  Driel-Gesztelyi}, {Valori}, \& {Dacie}}]{introduction_11}
{Yardley}, S.~L., {Green}, L.~M., {Williams}, D.~R., {et~al.} 2016, \apj, 827,
  151, \dodoi{10.3847/0004-637X/827/2/151}

\bibitem[{{Zhang} {et~al.}(2001){Zhang}, {Wang}, {Deng}, \&
  {Wu}}]{introduction_13}
{Zhang}, J., {Wang}, J., {Deng}, Y., \& {Wu}, D. 2001, \apjl, 548, L99,
  \dodoi{10.1086/318934}

\bibitem[{{Zirin}(1985)}]{introduction_5}
{Zirin}, H. 1985, Australian Journal of Physics, 38, 961,
  \dodoi{10.1071/PH850961}

\bibitem[{{Zuccarello} {et~al.}(2007){Zuccarello}, {Battiato}, {Contarino},
  {Romano}, \& {Spadaro}}]{zuca}
{Zuccarello}, F., {Battiato}, V., {Contarino}, L., {Romano}, P., \& {Spadaro},
  D. 2007, \aap, 468, 299, \dodoi{10.1051/0004-6361:20066556}

\end{thebibliography}



\end{document}